\documentclass[a4paper,10pt]{article}
\usepackage[english]{babel}
\usepackage{amsmath,amsfonts,verbatim,xkeyval,bm, amsthm, upgreek, amscd, wasysym, amssymb}
\usepackage{hyperref}
   \usepackage{graphicx}
\usepackage[dvipsnames,svgnames,x11names]{xcolor}
  \usepackage{ae,aecompl}
  \usepackage{cancel, ulem, enumitem}

\def\R{\mathbb{R}}

\def\E{\mathcal E}

\def\K{\mathcal K}

\let\e=\varepsilon

\newcommand{\be}{\begin{equation}}    
\newcommand{\ee}{\end{equation}}
\newcommand{\ba}{\begin{eqnarray}}
\newcommand{\ea}{\end{eqnarray}}
\newcommand{\bv}{\boldsymbol}
\newcommand\beq[1]{ \begin{equation}\label{#1} }
\newcommand{\eeq}{ \end{equation} }

\newcommand\beqa[1]{ \begin{eqnarray} \label{#1}}
\newcommand{\eeqa}{ \end{eqnarray} }
\newcommand{\beqano}{ \begin{eqnarray*} }
\newcommand{\eeqano}{ \end{eqnarray*} }

\newcommand{\nn}{\nonumber \\}

\newcommand{\sizedef}{
        \headheight=0pt                               
 	  \topmargin=-1.5cm \headsep=1.5cm              
        \oddsidemargin=-0.5cm \evensidemargin=-0.5cm  
        \textheight=22truecm \textwidth=16.5truecm    
 	  \setlength{\columnsep}{20pt}                  
  }

\sizedef

\begin{document}

\title{Perturbed isotropic oscillators with inversion symmetry}

\author{
  { Antonella Marchesiello$^1$}\\
  {Rita Mastroianni$^{2,3}$}\\
  { Giuseppe Pucacco$^4$}\\
   {\small$^1$ Department of Applied Mathematics, Faculty of Information Technology, Czech Technical University in Prague,}\\
  {\small Th\' akurova 9, 160 00 Prague}\\
  {\small$^2$  Advanced Concept Team, ESTEC, European Space Agency}\\
  {\small Keplerlaan 1, 2201 AZ Noordwijk, Netherlands}\\
  {\small$^3$ European Space Agency, Space Debris Office, Robert-Bosch-Str. 5, Darmstadt, Germany}\\
  {\small$^4$ Dipartimento di Fisica, Universit\`a
  degli Studi di Roma ``Tor Vergata'',}\\
  {\small via della Ricerca Scientifica 1, 00133 Roma}\\
  [1.2ex]
  {\small e-mails:
  {\tt  marchant@fit.cvut.cz, rita.mastroianni@esa.int}}\\
  {\small
  {\tt pucacco@roma2.infn.it}}
}

\date{}

\maketitle

\begin{abstract}
We consider the normal form  of a 1:1 detuned resonance that can be related, via a double averaging process, to the three body problem of celestial mechanics.  In view of the symmetry in the gravitational coupling between the planets, the obtained normal form exhibits inversion symmetry. We  provide an analytical prediction of  the possible  bifurcations in terms of the detunings and the other physical parameters of the system. We show that the bifurcation scenario is rich. It includes bifurcation sequences where the unstable orbit appearing after a saddle-node bifurcation can have a reversed pitchfork bifurcation, where it turns into stability with the birth of two unstable orbits, which eventually reconnect with one of the two quasi-normal modes. Furthermore, another scenario shows a standard pitchfork in which one of the quasi-normal modes becomes unstable, giving rise to a pair of stable periodic orbits.  By introducing coupling and asymmetry parameters, we provide an effective way to discern all the possible sequences of bifurcations.
    \end{abstract}

\section{Introduction}
The classical Hamiltonian model describing perturbations of the isotropic oscillator is one of the cornerstones for the understanding of non-linear dynamics. Explorations of near-integrability and transition to chaos on one side \cite{hehe,gunf} and of the structure of their integrable approximations
\cite{Ku,CHH} have played a fundamental role in the development of the theory. 

Several refinements have produced more sophisticated methods and tools to investigate bifurcation phenomena of invariant structures in systems characterised by a large variety of resonances and symmetries \cite{KE,Hlibro,SV}. Ideas from group theory and topological theory of dynamical systems provide powerful approach for a general understanding \cite{Br1:1,CB,VU11}.


Nonetheless, a complete survey of all possibilities is difficult to get due to the high dimensionality of the parameter space of the full model. Here we present the analysis of a case study emerged in the context of celestial mechanics \cite{RC23,RAG25}. In principle, this case is included in the pioneering work by Kummer \cite{Ku}: however, a detailed inspection of the phase-space structures determined by the relevant control parameters is still not available and can be useful in several applications.

The model is produced by a double average of the spatial restricted planetary 3-body problem. The Hamiltonian provided by the averaging is in the form of an almost isotropic perturbed 2D oscillator invariant under the inversion of the coordinate axes. The class of Hamiltonian normal forms obtained by resonant 1:1 normalization is  truncated at second order and expressed in terms of the Hopf invariants of the oscillator dynamics. The reduced system turns out to be given by a family of cylindrical conics constrained on the Hopf sphere and can be investigated by locating its critical points and assessing their nature. Back transforming to Cartesian variables, the phase-space structure of the integrable approximation of the original model can be determined \cite{RC23,RAG25}. Notice that this reduction is strictly valid in the regular set of initial conditions which avoid the singularities mentioned in \cite{PY2013}.

In agreement with Kummer \cite{Ku} we show that there are two types of critical points of the reduced system. For each of them we find  analytic formulas which predict the critical values of the Hopf radius (the {\it distinguished parameter}) that determine the bifurcations in terms of the {\it detuning} and the other control parameters. The appearance of two different types of critical points corresponds either to a pitchfork or to a saddle-node bifurcation. They are associated to the families of periodic orbits {\it in generic position} \cite{SV} of the original physical model and shape the overall phase-space structure. Several different sequences of these bifurcations occur in terms of the control parameters when varying the distinguished parameter.

By introducing a 3-dimensional parameter space (3DPS) defined by the {\it coupling} and {\it asymmetry parameters} \cite{STK,bif11}, we obtain a geometric method to classify the possible bifurcation sequences as they are determined by the control parameters and the detuning. Each type of bifurcations corresponds to a family of surfaces in the 3DPS: elliptical cones for the pitchforks and star-shaped cones for the saddle-nodes. A given system dynamics is represented by a straight line in the 3DPS parametrized by the distinguished parameter. Each crossing of this line with those surfaces produces a bifurcation. Examining all possible cases provides a classification scheme of the sequences.

The plan of the paper goes as follows: in Sect. \ref{sec:nota} we introduce notations and main mathematical objects; in Sect. \ref{sec:startH} we present the normal-form Hamiltonian and discuss its main properties; in Sect. \ref{sec:criticalPoints} we derive the critical points of the system on the Hopf sphere; in Sect. \ref{sec:bifsymmetric} we summarize the known results in the symmetric case of the $2:2$ resonance; in Sect. \ref{sec:bifurcations} we describe the bifurcations of the critical points  in the generic case of the inversion symmetry; in Sect. \ref{sec:catamap} we introduce the 3DPS for the classification of the bifurcation sequences; conclusions and final remarks follow in Sect. \ref{sec:conclusions}.

\section{Notations}\label{sec:nota}
Here we briefly recall the main notions adopted from \cite{VU11,bif11} and needed in the following.

\noindent
Let us consider a symplectic manifold with coordinates $P_1,P_2,Q_1,Q_2$ and fundamental 2-form \begin{equation} \label{2form}
\Omega=\sum_{k=1}^2 dP_k \wedge dQ_k\,.
\end{equation}
In the following, we shall focus on the {\it normal form} of the $1:1$ resonance, with special symmetry. Namely, let us consider the following transformations:
\begin{eqnarray}
\mathcal S_1&:&(P_1,P_2,Q_1,Q_2)\rightarrow(-P_1,P_2,-Q_1,Q_2)\,,\label{spatial_symmetry_1}\\
\mathcal S_2&:&(P_1,P_2,Q_1,Q_2)\rightarrow(P_1,-P_2,Q_1,-Q_2)\,,\label{spatial_symmetry_2}\\
\mathcal S_3 &:& (P_1,P_2,Q_1,Q_2)\rightarrow(P_2,P_1,Q_2,Q_1)\,,\label{scambioxy_simm}\\
\mathcal T&:&	  (P_1,P_2,Q_1,Q_2)\rightarrow(-P_1,-P_2,Q_1,Q_2)\,.\label{time_symmetry}
\end{eqnarray}
We are interested in the case in which the system is invariant  under the time reversal symmetry $\mathcal T$ and  the spatial {\it inversion symmetry}
\begin{equation}\label{inv_simm}
\mathcal S_1 \oplus \mathcal S_2: (P_1,P_2,Q_1,Q_2)\rightarrow(-P_1,-P_2,-Q_1,-Q_2).
\end{equation}

By construction, the normal form of the  $1:1$ resonance Poisson commutes with  the $1:1$ oscillator and it has polynomial dependence on its invariants  \cite{Murdock03}:
\begin{equation}\label{invariants}
\left\{\begin{array}{ll}
         I_0= & \frac12(P_1^2+P_2^2+Q_1^2+Q_2^2) =\E\\
         I_1= & P_1P_2+Q_1Q_2 \\
         I_2= & Q_1P_2-Q_2P_1 \\
         I_3= & \frac12(P_1^2-P_2^2+Q_1^2-Q_2^2)
       \end{array}
\right. \, ,
\end{equation}
 linked by the {\it syzygy}
\begin{equation} \label{rel_invariants}
I_1^2+I_2^2+I_3^2=I_0^2=\E^2 \,.
\end{equation}
Thus, the  dynamics of the normal form in these coordinates occur on the sphere $\mathcal S$ of equation \eqref{rel_invariants}. Their brackets 
\begin{equation} \label{rel_PB1}
\{I_a,I_0\}=0,\,\,\,\{I_a,I_b\}=2\e_{abc} I_c, \;\; a,b,c \in \{1,2,3\},
\end{equation}
determine the Poisson dynamics.

When working with resonant normal forms, it can be convenient to consider Cartesian variables in action-angles variables, given by
\be\label{Vaa}
Q_{\ell}=\sqrt{2J_{\ell}}\cos\phi_{\ell},\;\;\;\;\;P_{\ell}=\sqrt{2J_{\ell}}\sin\phi_{\ell}, \quad \ell=1,2 ,
\ee 
or  coordinates adapted to the resonance which, in case of the $1:1$ resonance, read \cite{Br1:1, Ver79}
\be \label{suber}
\left\{
  \begin{array}{l}
     J_1= J \\
     J_2=\E- J\\
     \psi=\phi_1-\phi_2\\
    \chi=\phi_2
  \end{array}
\right. \, .
\ee
Kummer's stereographic coordinates \cite{Ku} are given by
{\be\label{ccoord}
\left\{
  \begin{array}{l}
    x=\sqrt{2 J}\cos\psi \\
    y=\sqrt{2 J}\sin\psi
  \end{array}
\right.
\ee}
and will be useful to illustrate Poincar\'e surfaces. 

\section{The system}\label{sec:startH}

The motivation of our study arises from the Hamiltonian  related, via a double averaging process, to the three-body problem of celestial mechanics \cite{RC23}. 
The Cartesian variables $\bv{Q,P}$ are actually Poincar\'e variables related to eccentricities and peri-centers.
 The dynamics therefore describe the secular evolution of the orbits of two `planets' in the field of a `star'.
 
By neglecting higher order terms, after a diagonalization, the Hamiltonian is in the form of a slightly detuned isotropic oscillator perturbed by quartic terms.
 In view of the symmetry in the gravitational coupling between the planets it turns out that, in addition to the time reversal 
 symmetry, the system is invariant under the rotation of both arguments $\phi_{\ell}$ by $\pi$. After normalization with respect to the isotropic oscillator, the Hamiltonian is brought into the normal form \cite{RAG25}:
\begin{equation}\label{hamiltonian_inv}
\K(I_1,I_2,I_3; \E)= A I_1^2 + B I_3^2+ C I_1 I_3 + (D_1\E+\Delta_1) I_1 + (D_3 \E+ \Delta_3) I_3\,,
\end{equation}
where $I_j$ are the invariants of the isotropic oscillator (cf. \eqref{invariants}), and $A,B,C,D_1,D_3,\Delta_1,\Delta_3$ are control parameters. 
We remark that:
\begin{enumerate}
\item  $\Delta_1,\Delta_3$ act as `detunings', in the sense that when $\Delta_1=\Delta_3=0$ the unperturbed system would be at the exact $1:1$ resonance.
\item After normalization, the radius of the Hopf sphere $\E$ is a constant of motion for the Hamiltonian $\K$ (see \eqref{rel_PB1}), thus $\K$ defines an integrable system.  In the literature $\E$ is also referred as `distinguished parameter' \cite{Br1:1}, to discern its different nature from the other parameters ($A, B, \ldots$).
By neglecting higher order terms, $\E$ can also be interpreted as an approximation of the energy of the system \cite{bif11}. 
\item $\K$ does not depend on $I_2$ so that, in principle, we can reduce the investigation to the plane $I_1,I_3$. However, the 3D nature of the geometry remains fundamental as we see soon. If $C^2 > 4AB$, the level sets of the function $\K$ in $\R^3$ are a family of hyperbolic cylinders, otherwise they are a family of elliptical cylinders.
\item In view of the above symmetry, for generic values of the parameters, $\K$ is  not invariant separately under $\mathcal S_1,\mathcal S_2$, but only under the inversion symmetry \eqref{inv_simm}.
\end{enumerate}

In terms of action-angle like variables, the normal form reads
\begin{eqnarray}
{\mathcal K}_{AA} &=& 2(D_3 \E+ \Delta_3)J + 4BJ^2  + 2 \sqrt{J (\E-J)} (D_1\E+\Delta_1 + 2CJ) \cos \psi  + 2 A J (\E-J) \cos 2\psi. \label{hamiltonian_aa}
\end{eqnarray}
In terms of the stereographic coordinates we get (a part constant terms):
\begin{eqnarray}
{\mathcal K}_{SC} (x,y) &=& a_{20} x^2 + a_{02} y^2 + a_{40} x^4 + a_{04} y^4 + a_{22} x^2 y^2 \nn
&& + \left[D_1\E+\Delta_1 - C\left(\E - (x^2 + y^2)\right)\right] x \sqrt{2\E - (x^2 + y^2)}\label{hamiltonian_sc}
\end{eqnarray}
with the $a_{jk}$ combinations of $A,B,D_3,\Delta_3$.

Since, by a rotation, it is always possible to set $C=0$ in \eqref{hamiltonian_inv}, from now on we can assume $C=0$ and consider
\begin{equation}\label{hamiltonian_inv_noC}
\K(I_1,I_3; \E)= A I_1^2 + B I_3^2 + \delta_1 I_1 + \delta_3 I_3 \, ,
\end{equation}
setting
\begin{equation}\label{delta1-3}
\delta_1= (D_1 \E+ \Delta_1), \;\; \delta_3= (D_3 \E+ \Delta_3).
\end{equation}
Actually, the parameters appearing in \eqref{delta1-3} are a linear combination of those in \eqref{hamiltonian_inv}, but we keep the same symbols for sake of simplicity.

In terms of the invariants of the isotropic oscillator, the equations of motion provided by the Hamiltonian \eqref{hamiltonian_inv_noC}, by using  \eqref{rel_PB1}, then read

\begin{equation}
\label{eqMotion}
\begin{cases}
&\dot I_1=2 I_2 (2 B I_3 + \delta_3) \\
&\dot I_2= 2 \left( (\delta_1+ 2( A - B) I_1) I_3- \delta_3 I_1\right)  \\
&\dot I_3= -2 I_2 (2 A I_1 +\delta_1)
\end{cases}\, .
\end{equation}

We are interested in the study of the equilibria of the system  \eqref{eqMotion} on the sphere \eqref{rel_invariants} for $\E>0$.  In fact, for $\E=0$, the dynamics simply reduce to the equilibrium at the origin, i.e. the center of the sphere \eqref{rel_invariants}.
Let us first look at the values of the coefficients $A,B$ of \eqref{hamiltonian_inv_noC} for which the systems can exhibits non-isolated equilibria. We refer to these as degenerate cases.
\subsection{Degenerate case $A=B$}\label{sec:equidegAB}
We see that for $I_2=0$, the right hand side of the first and third equation of motion vanishes.  If $A=B$, the second equation is trivially satisfied for $\delta_1=\delta_3=0$. In this case the normal form simplifies to 
$$\mathcal K= A (I_1^2+ I_3^2) .$$
Therefore, at the energy level $\K= h=A\E^2$, the circle $I_1^2+ I_3^2=\E^2$ consists of non-isolated equilibria. 
\noindent
If $I_2\neq0$, we find the two (isolated) equilibria \eqref{EquiCenter}. As we show in Section \ref{sec:bifurcations}, these equilibria (when they exist) are stable, since $AB= A^2>0$. 
\subsection{Degenerate case $A B=0$}\label{sec:equidegABzero}
Excluding the trivial case in which all coefficients in the normal form vanish, looking for non-isolated equilibria  we see that we have other two possibilities:  $A=\delta_1=0$ or $B=\delta_3=0$.
In these cases, for $I_2\neq0$ the equations \eqref{eqMotion}  are solved by
$$I_3=-\frac{\delta_3}{2B}, \;\; I_1^2+ I_2^2= \E^2-\frac{\delta_3^2}{4 B^2} \quad (B\neq0)$$
and 
$$I_1=-\frac{\delta_1}{2A}, \;\; I_2^2+ I_3^2= \E^2-\frac{\delta_1^2}{4 A^2} \quad (A\neq0),$$
respectively. Thus, in both cases we can have a circle of equilibria on the sphere \eqref{rel_invariants}, provided that $\E^2-\frac{\delta_3^2}{4 B^2}\geq 0 $ and $\E^2-\frac{\delta_1^2}{4 A^2} \geq 0 $, respectively.

For $A=\delta_1=0$, $\mathcal K= B I_3^2+ \delta_3 I_3$. Therefore the level sets $\mathcal K=h$ on the  $(I_3, I_1)$ plane consist of a pair of straight lines parallel to the $I_1$ axis, except for $h=-\frac{\delta_3^2}{4B}$ when the two lines coincide at 
$I_3=-\frac{\delta_3}{2B}$. At $I_3=\pm \E$ the two lines are tangent to the circle, corresponding to these we have two equilibria. For $-\E<I_3< \E$ each energy level corresponds to a pair of planes  that cuts the sphere \eqref{rel_invariants} perpendicularly to the plane $I_2=0$, giving rise to a pair of periodic orbits around the equilibria on the sphere; thus, the two equilibria are stable. At 
$I_3=-\frac{\delta_3}{2B}$ these periodic orbits meet and break into a circle of equilibria. The situation is similar in the case $B=\delta_3=0$.
\subsection{Non-degenerate case $AB\neq0$ and $A\neq B$}
For $A B\neq0$ and $A\neq B$, all equilibria of the system are isolated. They
are given by $\mathcal Q\equiv (\mathcal Q_1, \mathcal Q_2, \mathcal Q_3)$, where
\begin{equation}\label{EquiCenter}
\mathcal Q_1= -\frac{ \delta_1}{2 A},\;\; \mathcal Q_3= -\frac{  \delta_3}{2  B} \qquad (A B\neq0)
\end{equation}  and $\mathcal P \equiv (\mathcal P_1, \mathcal P_2, \mathcal P_3)$ such that 
\begin{equation}\label{eqEquiOthers}
\mathcal P_2=0 \hbox{  and } (( \delta_1+ 2( A - B) \mathcal P_1) \mathcal P_3- \delta_3 \mathcal P_1)=0 \qquad (A \neq B) \, .
\end{equation}
Moreover, in order to be acceptable for the dynamics, $\mathcal Q$ and $\mathcal P$ have to lie on the sphere \eqref{rel_invariants}. By imposing the constraint \eqref{rel_invariants} and solving  the resulting equation in terms of $\E$, we are interested in the value of $\E$ that determines the appearance/disappearance of the equilibria  \eqref{EquiCenter}-\eqref{eqEquiOthers} on the sphere (see section \ref{sec:bifurcations}). There are always two solutions $\mathcal P_A$, $\mathcal P_B$ of \eqref{eqEquiOthers} (\cite{Ku}), corresponding to the so called quasi-normal (QN) modes (\cite{RC23}). However, depending on $\E$, another pair of equilibria solving \eqref{eqEquiOthers} might appear/disappear through  a  pitchfork bifurcation (if $D_1=\Delta_1=0$ or $D_3=\Delta_3=0$)  from $\mathcal P_A$ or $\mathcal P_B$ (that in this case are the usual normal modes \cite{bif11}) or a saddle-node bifurcation (see Section \ref{sec:SaddleNode}). Equilibria \eqref{EquiCenter} always appear/disappear  via a pitchfork bifurcation, as we will see in Section \ref{sec:pitchforks}. 

Notice that for $D_1 \Delta_1 <0$ and $D_3 \Delta_3 <0$ there exists positive values of $\E$ that satisfy the conditions $\delta_1=\delta_3=0$. However, as long as $A B\neq0$ and $A\neq B$, these values of $\E$ do not correspond to non-isolated equilibria. On the other hand, for $\delta_1$ or $\delta_3$ vanishing, the Hamiltonian \eqref{hamiltonian_inv_noC} exhibits additional symmetries; 
therefore  such values of $\E$ might determine a global change in the phase space structure of the system and correspond to so called `global bifurcations'. 

\subsection{Symmetries}\label{sec:symmetries0} 
Let us discuss here in more detail the symmetries of the system. In this work we are interested in the case when the system is invariant under the inversion symmetry \eqref{inv_simm}, hereafter referred to as Case IV). However, for special values of the parameters, the system could exhibit additional symmetry. This is why we also refer to the case of inversion symmetry as the `generic case'. Namely, depending on the values of the parameters, 
we have the following cases:
\begin{description}
\item[Case I)]  $D_1=D_3=\Delta_1=\Delta_3=0$: 
`symmetric case'.  In this case   the system is symmetric under all symmetries $\mathcal S_j$, c.f. \eqref{spatial_symmetry_1}, \eqref{spatial_symmetry_2} and \eqref{scambioxy_simm}.
As a consequence, $\delta_1,\delta_3$ are identically zero  and the normal form \eqref{hamiltonian_inv_noC} is symmetric under 
\begin{equation}\label{symmR12}
\mathcal R_1 :\,\, I_1\longrightarrow - I_1, \;\;\; \mathcal R_2:\,\, I_2\longrightarrow -I_2
\end{equation}
and
\begin{equation}\label{symmR3}
\mathcal R_3 :\,\, I_3\longrightarrow - I_3.
\end{equation}
In the non-degenerate cases, the system has six equilibria, that can be easily found from \eqref{EquiCenter} and \eqref{eqEquiOthers}. Each equilibria lies  at one of the poles of the sphere; namely, they  are given by
\begin{description}
\item a) 
\be I_1=\pm\E, \;\;  I_2=I_3=0\,,\label{equia}
\ee
\item b) 
\be I_2=\pm\E, \;\;  I_1=I_3=0 \,,\label{equib}
\ee
\item c) 
\be I_3=\pm\E,\;\; I_1=I_2=0\,.\label{equic}\ee
\end{description} 

Examples of the dynamics in Case I are provided in Fig.\ref{fig:ellipses00} when $AB>0$ and in Fig.\ref{fig:hyperbolas00} when $AB<0$.
\begin{figure}[!]
\begin{center}
\includegraphics[scale=.4]{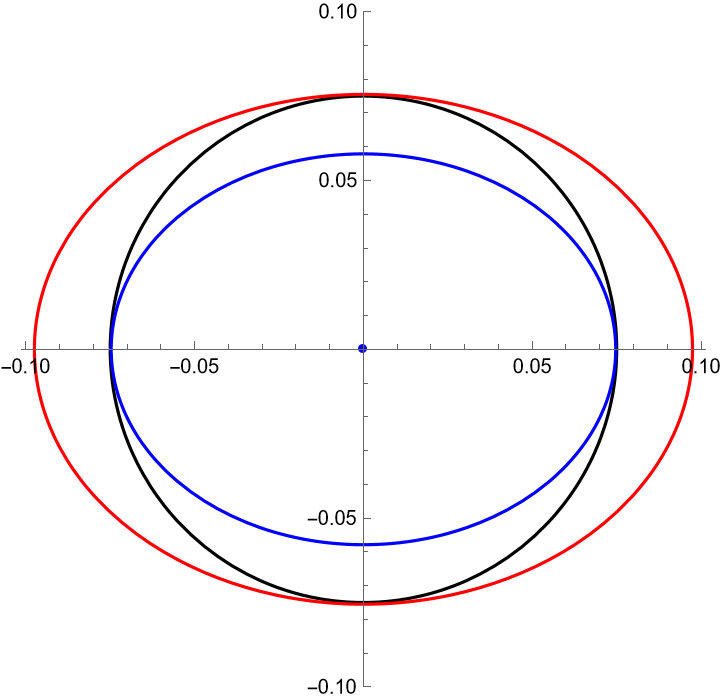}
\quad
\includegraphics[scale=.33]{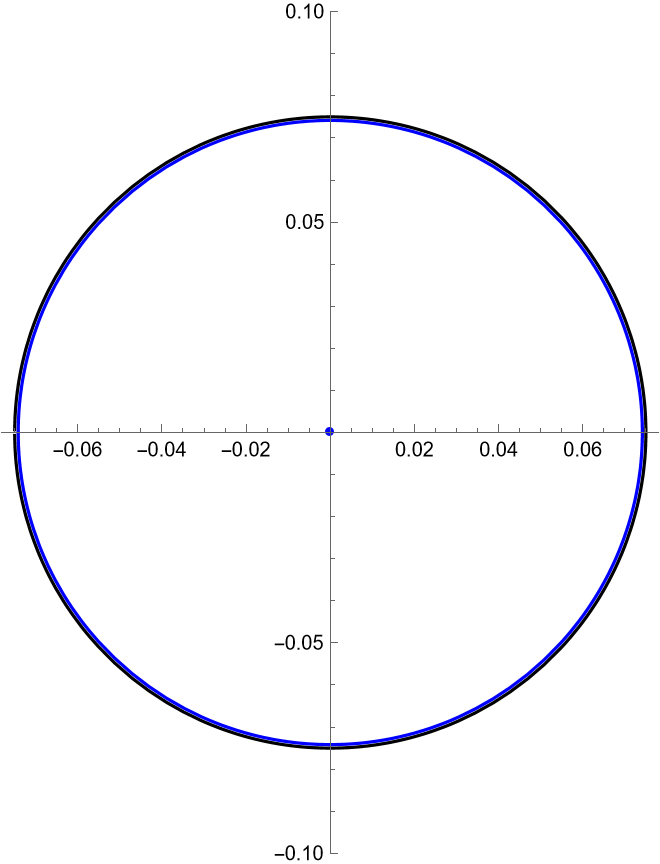}
\quad
\includegraphics[scale=.33]{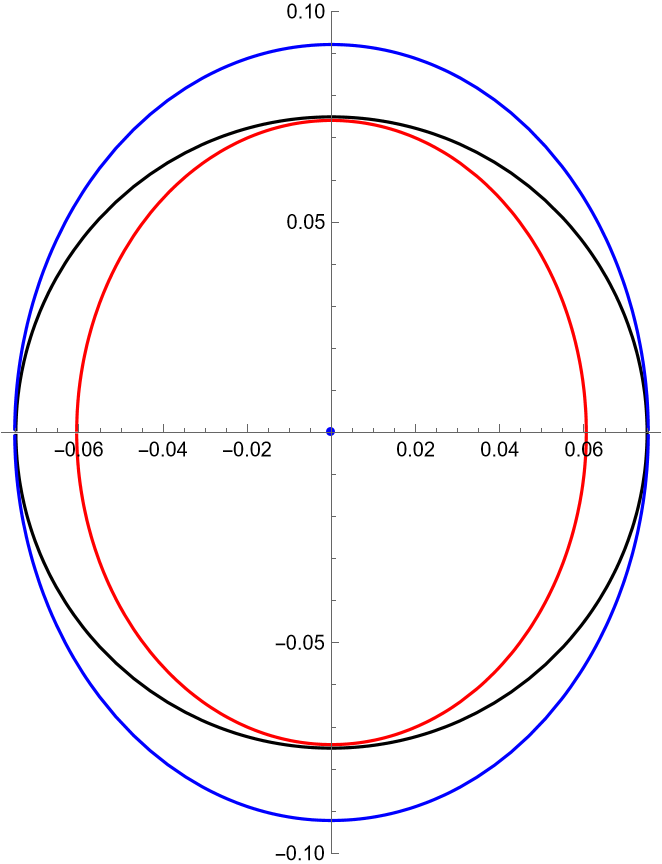}
\caption{Section of the phase space sphere (in black) and the level sets of the normal form corresponding to equilibria on the plane $I_2=0$ for $D_1=D_2=\Delta_1=\Delta_3=0$, $AB>0$, $\E=0.075$. Left: $A=0.6$, $B=1$. Center: $A=B=0.6$. Right: $A=0.6$, $B=0.4$. The tangency points between the sphere (in black) with the blue and red ellipses correspond to equilibria a) and c), respectively. The blue point at the origin represents equilibria b).}\label{fig:ellipses00}
\end{center}
\end{figure}
\begin{figure}[!]
\begin{center}
\includegraphics[scale=.4]{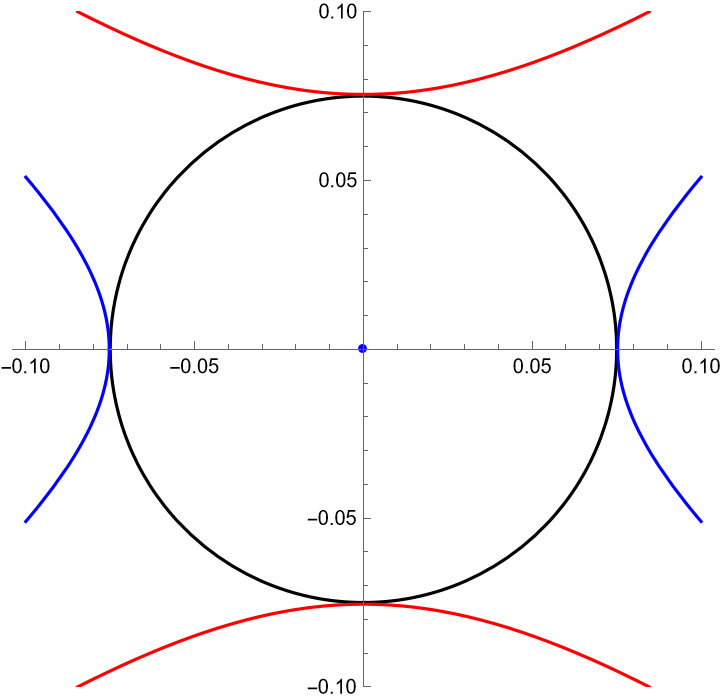}
\quad
\includegraphics[scale=.4]{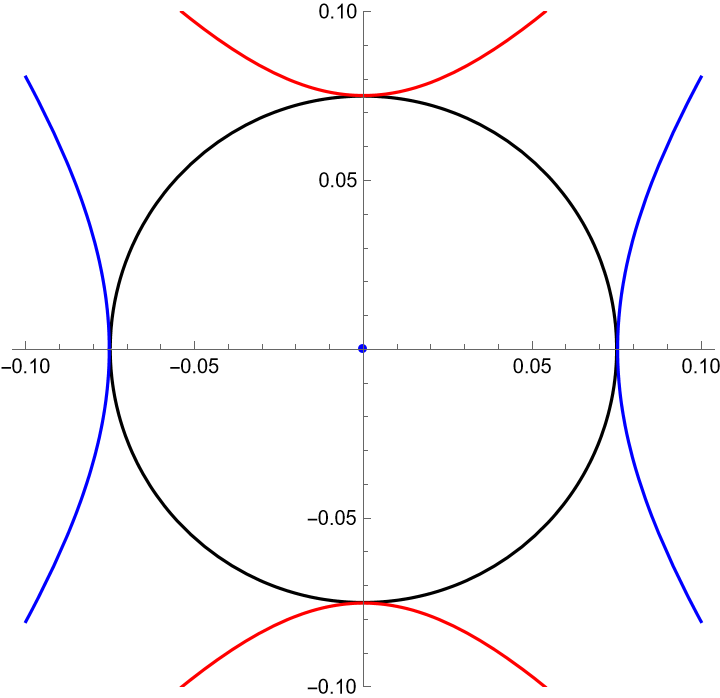}
\caption{Section of the phase space sphere  (in black) and the level sets of the normal form corresponding to equilibria on the plane $I_2=0$ for $D_1=D_2=\Delta_1=\Delta_3=0$, $AB<0$ and $\E=0.075$. Left: $A=-0.6$, $B=1$. Right: $A=0.6$, $B=-0.4$. The tangency points between the sphere (in black) with the blue and red hyperbolas correspond to equilibria a) and c), respectively. The blue point at the origin represents equilibria b).}\label{fig:hyperbolas00}
\end{center}
\end{figure}

\item[Case II)] $D_1=\Delta_1=0$: the `$2:2$ resonant case'. In this case $\delta_1\equiv0$ for every value of $\E$, thus the system is symmetric under both $\mathcal S_1$ and $\mathcal{S}_2$, c.f. \eqref{spatial_symmetry_1}-\eqref{spatial_symmetry_2}, and therefore the normal form \eqref{hamiltonian_inv_noC} is symmetric under 
\eqref{symmR12}. In this case we also speak of $\mathbb Z_2\times \mathbb Z_2$ symmetry and the normal form is related to the $2:2$ resonance \cite{bif11}.
\item[Case III)] $D_3=\Delta_3=0$:  the `rotated $2:2$ resonant case'. In this case $\delta_3\equiv0$, and thus the system exhibits symmetry under \eqref{symmR3}. 
By a rotation of $\frac{\pi}{2}$ around $I_2$ composed with a rotation of $\pi$ around $I_3$, we can exchange $I_1$ with $I_3$, thus we can reduce this case to the $D_1=\Delta_1=0$ case.

At the special cases II)-III) two poles of the sphere are equilibria, for any value of $\E$ (respectively, the est-west poles $I_1=I_2=0$, $I_3=\pm \E$ and the north-south poles $I_1=\pm \E$, $I_2=I_3=0$). These are the (QN)-modes that due to the symmetries are moved to the poles of the sphere, becoming  the standard {\it normal modes} of the $2:2$ resonance (\cite{bif11}).  However, by varying $\E$, other equilibria can appear/disappear, always via {\it{pitchfork bifurcations}} (\cite{bif11}). These bifurcating equilibria correspond to the other equilibria that would be at the poles in the  symmetric case I).

\item[Case IV)]$D_1^2+ \Delta_1^2\neq0$ and $D_3^2+ \Delta_3^2\neq0$: `the inversion symmetry case'. This is the generic case in which $\delta_1\delta_3\not\equiv0$, except at the values of $\E$ solution of $\delta_j=0$ . Thus, the original Hamiltonian  is invariant only under $\mathcal S_1\oplus \mathcal S_2$ and the normal form \eqref{hamiltonian_inv_noC} is symmetric only w.r.t.  $\mathcal R_2$.
This is the case of interest of this paper. As we shall see in Section \ref{sec:bifurcations}, for values of the parameters $D_j,\Delta_j$ (in the normal form) that break the symmetry with respect to both $\mathcal R_1$ and $\mathcal R_3$, one of the pitchfork bifurcations of  Cases II), III) becomes a {\it saddle-node} bifurcation. More precisely, since the  $\mathcal R_2$ symmetry is preserved, the pitchfork bifurcation that survives is the one that  involves the equilibria off the plane $I_2=0$.

Let us notice that at the critical values $\E=\E_{gb,j}$ for which $\delta_j=0$ (if any) the system passes through the symmetries $\mathcal R_j$. In analogy with Cases II)-III), this could be associated to a  {\it global bifurcation} (\cite{bif11}). This kind of bifurcation is so called because it determines a global change in the phase space structure of the system, though it is not associated to the appearance/disappearance of equilibria (see e.g. Figures \ref{fig:nosymmgb}-\ref{fig:nosymmgbps}). However, an investigation of  the global bifurcation is out of the scope of this paper. We shall focus in the following on  the local bifurcations in Case IV) (such as the pitchfork and saddle-node bifurcations) aiming at finding the conditions for the appearance/disappearance of  equilibria for the system. 
\end{description}

\section{The geometric method: critical points}\label{sec:criticalPoints} 

The equilibria of the system can also be found by looking for the critical points (CP) of the system, that are the contacts between the level sets $\mathcal K= h$ (where the Hamiltonian $\K$ is described in Eq.~\eqref{hamiltonian_inv_noC})
and the reduced phase-space \eqref{rel_invariants} (see Kummer \cite{Ku} and Cushman \cite{CHH}). This provides a geometric point of view  that will be useful in the following to understand the bifurcations of the system.

Notice that the level sets 
\be\label{conical}
\K=h
\ee  correspond to families of elliptical cylinders for $AB>0$ and hyperbolic cylinders for $AB<0$. 

We define the function
$$ {\cal F} (I_1,I_2,I_3; \E) = {\cal K} - \lambda {\cal G} \, ,$$
where $\lambda$ is a Lagrange multiplier and 
$$ {\cal G} = I_1^2+I_2^2+I_3^2-\E^2 \, ,$$
and look for the solutions of 
$$ \nabla {\cal F} = 0, $$
namely
\be\label{crit_eq}
\left(\begin{array}{ccc}
   2 (A-\lambda) & 0 & 0 \\
   0 & - 2 \lambda & 0 \\
   0 & 0 & 2 (B-\lambda) \end{array}\right) \;\; 
   \left(\begin{array}{c}
   I_1 \\ I_2 \\ I_3 \end{array} \right) \;\; = \;\;
   -\left(\begin{array}{c}
   \delta_1 \\ 0 \\ \delta_3 \end{array} \right)\, ,
\ee
with the constraint
\be\label{Gconstr} {\cal G}  (\lambda,\E) = 0, \ee
for $\delta_1$, $\delta_3$ as defined by Eq.~\eqref{delta1-3}. Following Kummer \cite{Ku},
the solutions of \eqref{crit_eq} are of two kinds, as illustrated in the following.

\subsection{Critical points of the first kind}\label{sec:CP1}
The equilibria  associated to the critical points of the first kind (CP1) occur at the tangency points between the level sets \eqref{conical}
of the normal form \eqref{hamiltonian_inv_noC}  and the sphere \eqref{rel_invariants}. Since the normal form $\mathcal K$ does not depend on $I_2$ and the phase space is a surface of revolution, a tangency point always lies on the plane $I_2=0$. In fact, the critical points of the first kind are the solutions of \eqref{crit_eq}-\eqref{Gconstr} at $I_2=0$, corresponding to the equilibria $\mathcal P$  which solve Eq.~\eqref{eqEquiOthers}). As we observed in the previous Section, if $\delta_i=0$ the system passes through special symmetries. In these cases equation \eqref{eqEquiOthers} or correspondingly \eqref{crit_eq}-\eqref{Gconstr}, can be easily solved. By excluding the degenerate case $A=B$, at which there can be non-isolated solutions (see Section \ref{sec:equidegAB}), we have:
\begin{description}
\item[Case I)] If $\delta_1=\delta_3=0$ we find a pair of equilibria for $\lambda=A$, $ I_2=I_3=0$ and a pair  for $\lambda=B$, $ I_1=I_2=0$, that are equilibria a) and c) at the poles of the sphere (see Eqs.~\eqref{equia},~\eqref{equic}). 
\item[Case II)] For $\delta_1=0$, $\delta_3\neq0$ by solving \eqref{crit_eq}-\eqref{Gconstr} for $I_2=0$ we find 
\begin{description}
\item i) \begin{equation}\label{SolCP1delta1zero-B}
 \lambda=B\pm \frac{\delta_3}{2\E},\;\; I_3=\pm\E,\;\;I_1=I_2=0\,,
\end{equation}
\item ii) \begin{equation}\label{SolCP1delta1zero-A}
\lambda=A, \;\; I_3 = \frac{\delta_3}{2(A-B)}, \;\; I_1^2=\E^2-\frac{\delta_3^2}{4(A-B)^2} \;\; I_2=0\,.
\end{equation}
\end{description} 
\item[Case III)] Similarly, for $\delta_1\neq 0$, $\delta_3=0$ we find
\begin{description}
\item i) \begin{equation}\label{SolCP1delta3zero-A}
 \lambda=A\pm\frac{\delta_1}{2\E}, \;\;I_1=\pm \E,\;\; I_3=I_2=0\,,
\end{equation}
\item ii) \begin{equation}\label{SolCP1delta3zero-B}
\lambda=B, \;\; I_1= -\frac{\delta_1}{2(A-B)}, \;\; I_3^2= \E^2- \frac{\delta_1^2}{4(A-B)^2},\;\; I_2=0\,.
\end{equation} 
\end{description}
\end{description}
\begin{description}
\item[Case IV)] Finally, we have the case
in which $\delta_1\delta_3\neq0$. The CP1 are determined by the solutions of  Eqs.~\eqref{crit_eq}
\begin{equation}
\label{eqCP1}
\begin{split}
I_1^S (\lambda,\E)&=- \frac{ \delta_1}{2(A-\lambda)} \\
I_2^S (\lambda,\E)&=   0       \\    
I_3^S (\lambda,\E)&= -\frac{\delta_3}{2(B-\lambda)},    
\end{split}
\end{equation}
constrained by \eqref{Gconstr} to satisfy
\be\label{eqEquilambdaOriginal}
 {\cal G} (\lambda,\E) = (I_1^S)^2 + (I_3^S)^2 - \E^2 = 0 \, ,
 \ee
corresponding  to the solutions of \eqref{eqEquiOthers}. In  this case  we cannot find explicit analytic expression for the equilibria.
 \end{description}

\subsection{Critical points of the second kind}\label{sec:CP2}
The critical points of the second kind (CP2) are the equilibria that occur for $\lambda=0$.  

At the degenerate case $A=\delta_1=0$, the system \eqref{crit_eq} is solved at $\lambda=0$ by $I_3=-\frac{\delta_3}{2B}$, while both $I_1$ and $I_2$ are constraint only by \eqref{Gconstr}, thus we obtain a circle of equilibria (see Section \ref{sec:equidegABzero}). Similarly, for $B=\delta_3=0$.

If $AB\neq0$ there can be isolated solutions of \eqref{crit_eq} 
with ${\cal G} (0,\E) = 0$, given by
\begin{itemize} 
\item[iii)] 
 \be\label{CP2nd}
\left(I_1^S (0,\E),\pm \sqrt{\E^2-(I_1^S (0,\E))^2-(I_3^S (0,\E))^2}, I_3^S (0,\E) \right), \ee
\end{itemize}
where $I_1^S (0,\E)$ and $I_3^S (0,\E)$ correspond to $\mathcal Q_1$ and $\mathcal Q_3$, respectively (see Eq.~\eqref{EquiCenter}). This critical point turns out to be the center of each family of curves given by 
\eqref{hamiltonian_inv} (see Eq. 4.9 in~\cite{Ku}). 

Correspondingly to the CP2~\eqref{CP2nd}, at 
\begin{equation}
\label{def.hbar}
h= \bar h\equiv- \frac{A \delta_3^2+B \delta_1^2}{4 A B}
\end{equation}
the conical surface \eqref{conical} degenerates to a straight line parallel to the $I_2$ axis and passing through the CP \eqref{CP2nd}. Thus, if this is the case on the section $I_2=0$ of the phase space, at $h=\bar h$, \eqref{conical} coincides with the center of a family of ellipses ($AB>0$) or hyperbolas ($AB<0$).
Geometrically, the appearance of the CP2 is due to the centers of the ellipses or hyperbolas {\it entering or leaving the circle} 
 \begin{equation}\label{circumference}
I_1^2+ I_3^2=\E^2.
 \end{equation} 
In the three dimensional space, we have then a straight line parallel to the $I_2$-axis tangent to the sphere at $I_2=0$ and, when the line enters into the sphere, it {\it pierces} it in the two points \eqref{CP2nd}. Notice that for $\delta_1=\delta_3=0$ these corresponds to a pair of the equilibria at the poles $I_1=I_3=0$ of the sphere.
 
 \section{Pitchfork bifurcations  in the symmetric cases II)-III)}\label{sec:bifsymmetric}
 The equilibria ii)-iii) of cases II)-III) do not always exists. Namely we need $I_1^2$ and $I_3^2$ positive for Case II)-ii) and Case III-ii), respectively. For equilibria iii) the square root in \eqref{CP2nd} must be well defined. These conditions determine the bifurcations of the equilibria in the symmetric cases II)-III). They can be derived from \cite{bif11}. For sake of clarity, we summarize the results in the following in the non-degenerate cases  $A B \neq 0$ and $A\neq B$.  The bifurcations of the general case IV), that is the one of interest in this work, will be studied in the next Section \ref{sec:bifurcations}.
 
 \subsection{Case II)}\label{sec:2:2}
 In  Case II), namely $D_1=\Delta_1=0$, equilbria i) are CP1, corresponding to the south and north  pole of the sphere \eqref{rel_invariants}, namely $(0,0,-\E)$ and $(0,0,\E)$, respectively. For the original system in two degrees of freedom, these are the normal modes $J_1=0$, $J_2=\E$ (NM-) and $J_1=\E$, $J_2=0$ (NM+), respectively.  The remaining  equilibria ii)-iii) (CP1 and CP2, respectively) correspond to the so called inclined orbits ($ I_2=0, I_3=\frac{\delta_3}{2(A-B)}$) and loop orbits ($ I_1=0,I_3=-\frac{\delta_3}{2B}$), respectively. Notice that loops orbits correspond to equilibria \eqref{EquiCenter}, while equilibria i), ii) correspond to solutions of \eqref{eqEquiOthers} in case $\delta_1=0$.
Depending on the parameters, equilibria ii)-iii) can appear/disappear from north or south pole of the sphere through pitchfork bifurcations (\cite{bif11}). The bifurcations occur when the equilibria ii) or iii) arise from i), i.e., respectively, for the values of $\E$ s.t. $\delta_3/(2(A-B))=\pm\E$ or $-\delta_3/(2B)= \pm\E\,$. For $D_1=\Delta_1=0$, the corresponding critical values of $\E$ that determine the bifurcations of equilibria ii) and iii) from/off the normal mode NM$\pm$ are given by
\begin{equation}\label{enec-inclined-2:2}
\E_{ i, 1\pm}=  \frac{\Delta_3}{-D_3\pm 2 (A-B)}
\end{equation}
and
\begin{equation}\label{enec-loop-2:2}
\E_{\ell, 1\pm}= - \frac{\Delta_3}{ D_3\pm 2 B } \, ,
\end{equation}
respectively (see \cite{bif11}). 
Equilibria ii) are elliptic for $A(A-B)>0$ and hyperbolic for $A(A-B)<0$, while equilibria iii) are elliptic for $AB>0$ and hyperbolic for $AB<0$. 

\subsection{Case III)}
In this case, namely $D_3=\Delta_3=0$, the system acquires symmetry under \eqref{scambioxy_simm}. As already observed, we just have a ``rotated'' $2:2$ resonance. More precisely, modulo a rotation that brings the eastern and western poles of the sphere onto the north and south poles, the cases $\delta_1\equiv0$ and $\delta_3\equiv0$ are the same.
By renaming $A$ with $B$ and vice-versa, and replacing $D_3$ by $D_1$ and $\Delta_3$ by $\Delta_1$, one can easily find the equilibria and their stability properties from the analysis of the $2:2$ resonance. Let us give the result here for sake of clarity.
The isolated equilibria of the system are given by i), ii), iii) as in \eqref{SolCP1delta3zero-A}, \eqref{SolCP1delta3zero-B} and \eqref{CP2nd}, respectively. Explicitly, iii) in this case reads
\be \hbox{iii)}\;\; I_1=-\frac{ \delta_1}{2 A},\;\; I_3=0, \;\; I_2^2=\E^2-\frac{\delta_1^2}{4A^2}\,. 
\ee 
In case i), we have the eastern and western  poles of the sphere $(\pm \E, 0,0)$ on the plane $I_2=0$. For the original system in two degrees of freedom, these correspond to   $ 4 J_1 J_2= \E^2$, i.e. $J_1=J_2=\frac{\E}{2}$. The bifurcation scenario is similar to the one of the previous case.  For completeness, let us provide the critical values of $\E$ that determines the bifurcations also in this case.

For $D_3=\Delta_3=0$, in case ii) we have two inclined orbits for the original system that bifurcate when 
$$\E^2- \frac{\delta_1^2}{4(A-B)^2}=0,$$
that is  at the critical value
\begin{equation}\label{E1scambio}
\E_{i,2\pm}= \frac{\Delta_1}{-D_1\pm 2( A- B)},
\end{equation}
acceptable if not negative. Both equilibria are unstable for $B (A-B)>0$, otherwise stable for $B (A-B)<0$. In case iii), by solving $$\E^2- \frac{\delta_1^2}{4A^2}=0,$$
we find  the critical value
\begin{equation}\label{E2scambio}
\E_{\ell,2\pm}=- \frac{\Delta_1}{D_1\pm 2 A},
\end{equation}
that determines the bifurcation of the equilibria. 
These equilibria are stable if $AB >0$, unstable if  $AB <0$. Moreover, equilibria i) change their stability at the critical values $\E$ described in Eq.~\eqref{E2scambio} via a pitchfork bifurcation. 

\begin{figure}[htb!]
\begin{center}
\includegraphics[scale=.4]{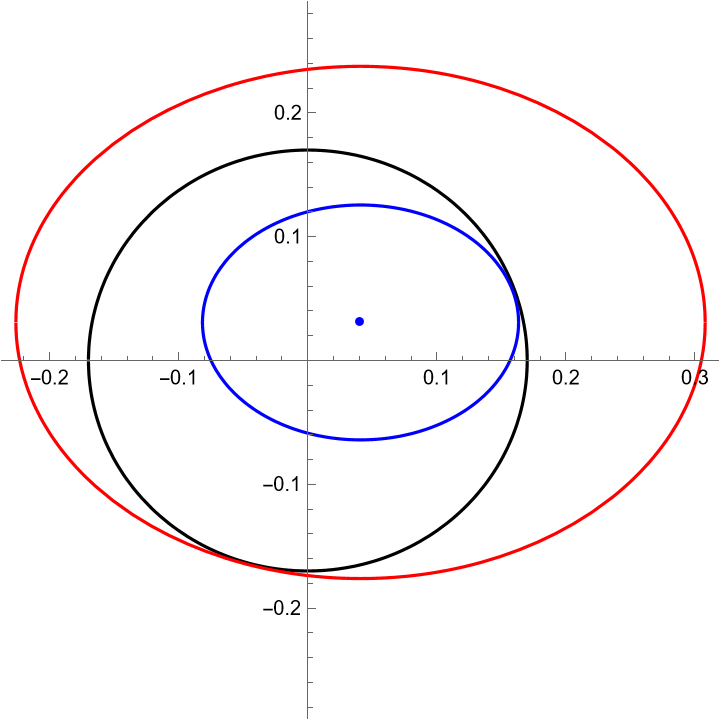}
\quad
\includegraphics[scale=.4]{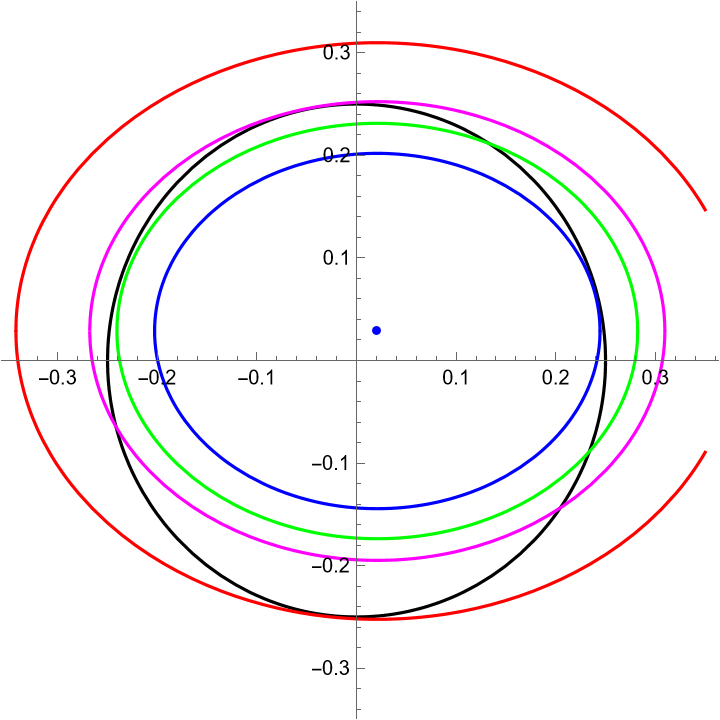}
\\
\includegraphics[scale=.4]{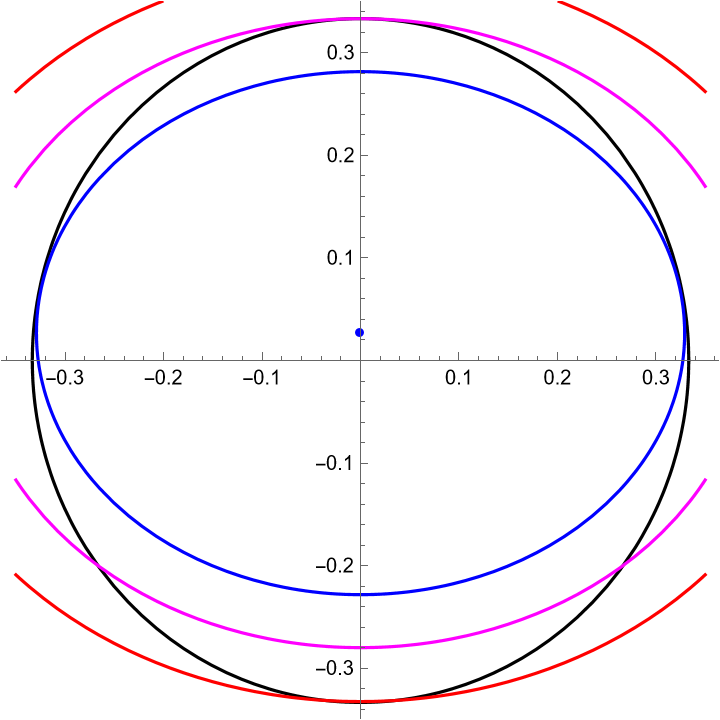}
\quad
\includegraphics[scale=.4]{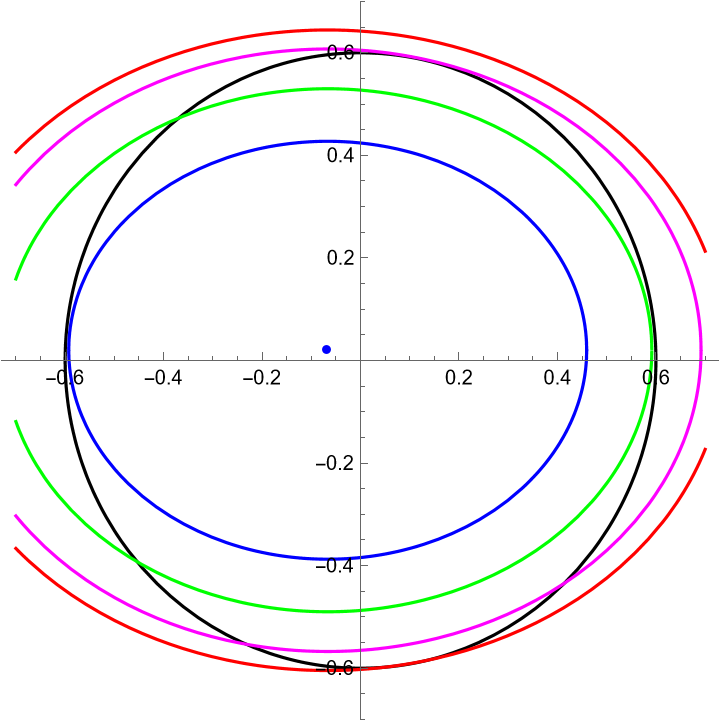}
\end{center}
\caption{Section of the sphere \eqref{rel_invariants} (in black) and the level sets  \eqref{conical}  on the plane $(I_1, I_3)$ for $ A =0.6, B = 1, D_1= 0.3,  \Delta_1= -0.1$, $D_3=0.05$, $\Delta_3=-0.07$.   Two equilibria are  given by the center of the ellipses $\K =\bar h\,$, i.e. $I_1=-\delta_1/(2A), I_2, I_3=-\delta_3/(2B)$ (the blue point, where $\nabla_{(I_1, I_2, I_3)} \K =0\,$) and appear after a pitchfork bifurcation, see top, first panel. The other four equilibria can be found numerically by Eqs. $I_2=0\,$, $I_3=\delta_3 I_1/\left((2(A-B) I_1 + \delta_1)\right)$ and $I_1^2 + I_3^2=\E^2\,$.  The pitchfork bifurcation of  Case III) (that involves the equilibria that lye on the plane $I_2=0$) is replaced by a saddle-node bifurcation, see top, second panel. 
We also observe a global bifurcation at $\E=\E_{gb,1}=0.\bar 3$, when the stable-unstable manifold of the two unstable equilibria (the green and the blue ellipses) coincide (see the blue ellipse, first bottom panel). (From left to right, top: $\E=0.17$, $\E=0.25$, bottom:  $\E=0.\bar 3$, $\E=0.6$). In Figure \ref{fig:nosymmgbps} we see the corresponding Poincar\'e surfaces in the stereographic coordinates \eqref{ccoord}.}
\label{fig:nosymmgb}
\end{figure}

\begin{figure}[htb!]
\begin{center}
\includegraphics[scale=.4]{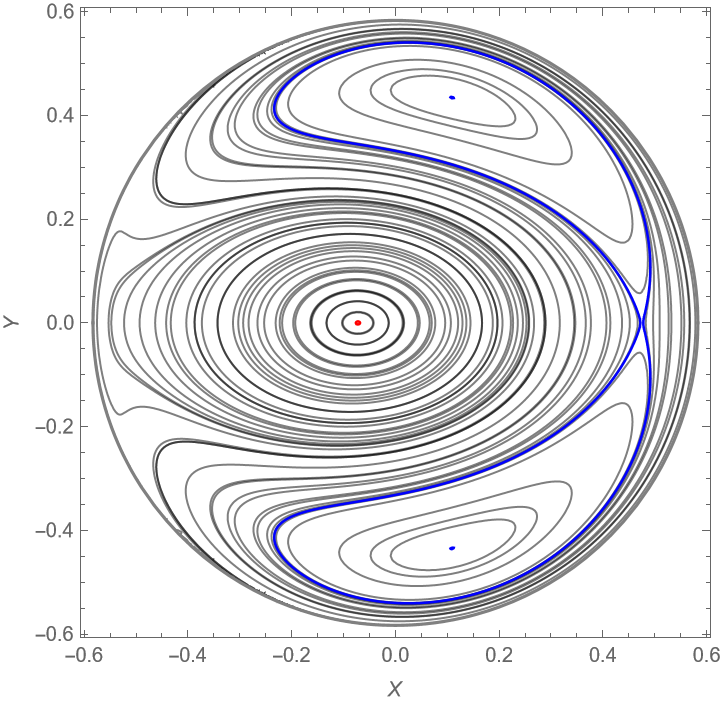}
\quad
\includegraphics[scale=.4]{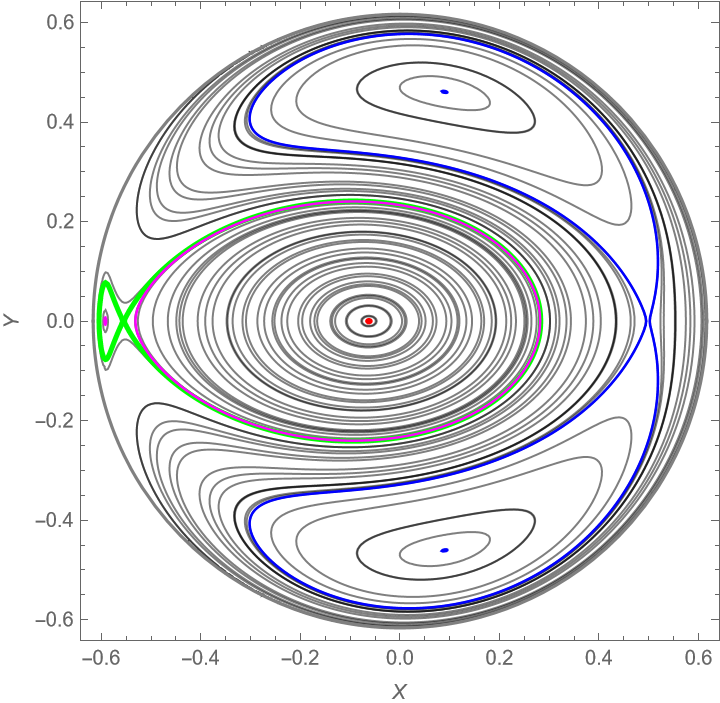}\\
\includegraphics[scale=.4]{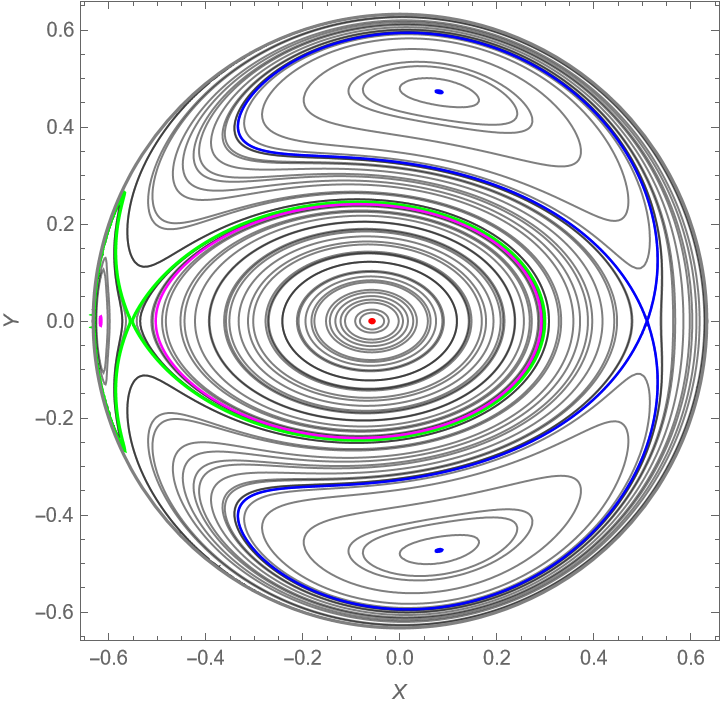}
\quad
\includegraphics[scale=.4]{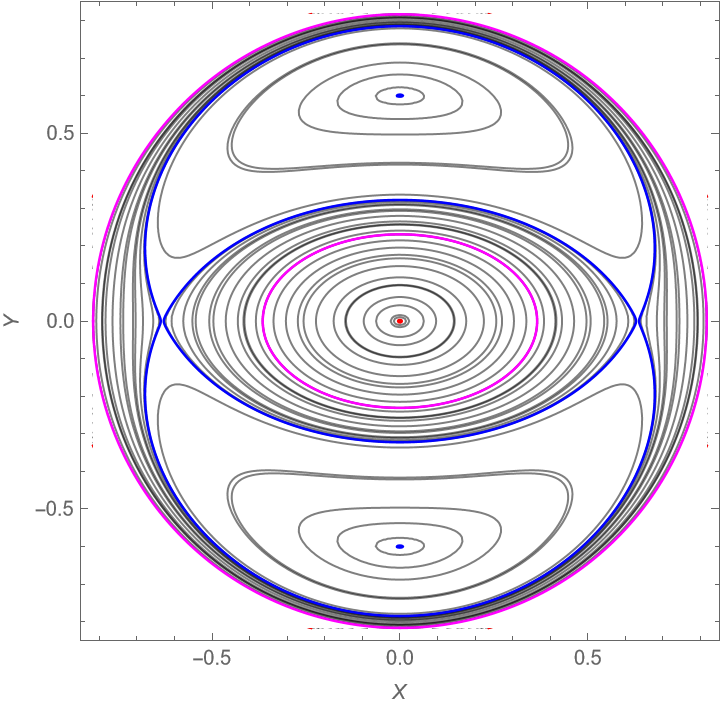}
\quad
\includegraphics[scale=.4]{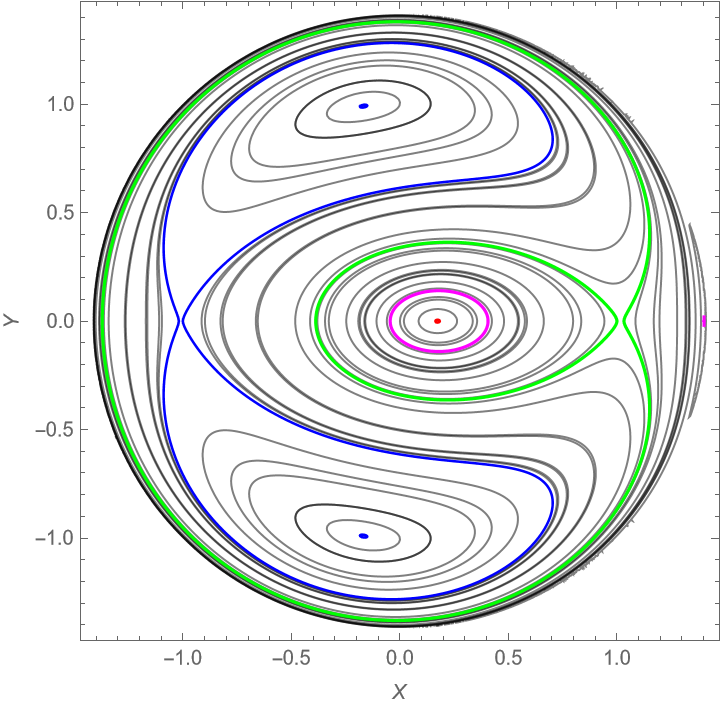}
\end{center}
\caption{Surfaces of section in the $(x,y)$ stereographic coordinates defined in \eqref{ccoord} corresponding to the cases of Fig.~\ref{fig:nosymmgb}. 
Upper left: we are after the pitchfork bifurcation ($\E=0.17$). Upper right: after breaking the symmetry of Case III), the pitchfork bifurcation on the plane $I_2=0$ has become a saddle-node ($\E=0.19$). 
Bottom: we see how the unstable  equilibria that occur at $y=0$ exchange their roles in the bifurcation sequence at the global bifurcation. In the first frame, we see that, before the global bifurcation, the pitchfork bifurcation has happened from the unstable equilibrium to the right (in blue). However, after the global bifurcation (third frame), the corresponding stable equilibria are enclosed by the stable-unstable manifold (in blue) of the equilibrium to the left. Similarly the homoclinic orbit (in green) of the unstable equilibrium  to the left encloses the equilibrium at the origin in the left frame. However, in the right frame, it is the homoclinic orbit of the equilibrium to the right that surrounds the origin (in green). At the global bifurcation, the stable-unstable manifold of the unstable equilibria coincide (middle frame, in blue). (The energies at bottom left/center/right frames are respectively: $\E=0.2$, $\E=0.\bar 3$, $\E=0.6$.)} 
\label{fig:nosymmgbps}
\end{figure}

\section{Local bifurcations of the inversion symmetry}\label{sec:bifurcations}
We aim here at understanding which kind of bifurcations are related to the solutions of equations \eqref{crit_eq}-\eqref{Gconstr} in the symmetric case IV) of the inversion symmetry. As anticipated in Section \ref{sec:symmetries0}, we will see that the CP2 appear/disappear via pitchfork bifurcations as in the symmetric cases II) and III), instead the CP1 appear/disappear through saddle-node bifurcations.

\subsection{Pitchfork bifurcations of the CP2}\label{sec:pitchforks}
As we already observed,  the appearance of the CP2 is due to the centers of the ellipses or hyperbolas {\it entering or leaving the circle} \eqref{circumference}
on the plane $I_2=0$. 
Therefore the bifurcation threshold is given by the positive root $\E_{CP2}$ of
\begin{equation}\label{eqBifCP2}
 (I_1^S (0,\E))^2+(I_3^S (0,\E))^2= \E^2.
 \end{equation}
We can simplify the problem and solve this condition for $I_2=0$, i.e. we find the value $\E_{CP2}$ at which the center of the conic section corresponding to \eqref{conical} lies on the circumference \eqref{circumference}.
For $AB\neq0$, we find two solutions when $B^2D_1^2+A^2(-4B^2+D_3^2)\neq0$, given by
\begin{equation}\label{sol.CP2}
\begin{split}
&\E_{CP2,\pm}=-\frac{B^2 D_1\Delta_1+A^2D_3\Delta_3\,\pm\,A B \sqrt{4 B^2\Delta_1^2-(D_3\Delta_1+2A\Delta_3-D_1\Delta_3)(D_3\Delta_1-(2A+D_1)\Delta_3)}}{B^2D_1^2+A^2(-4B^2+D_3^2)}\, .
\end{split}
\end{equation}
acceptable when well defined and not negative.

By the Hessian matrix of \eqref{hamiltonian_inv_noC} it can be proven that they are elliptic fixed points for $AB>0$ (ellipses) and hyperbolic fixed points for $AB<0$ (hyperbolas) (see Figs. \ref{fig:nosymmgb} and \ref{fig:nosymmgbps} for a case with ellipses and \ref{fig:hyperbolas1}, \ref{fig:hyperbolas2} for a case with hyperbolas.). Since  two equilibria appear simultaneously and have the same stability/instability, at the critical value \eqref{sol.CP2}, necessarily we have a pitchfork bifurcation. This is in agreement with the Poincar\'e-Hopf Index Theorem \cite{PoiHopfBook}, since the Euler characterisic of a sphere is two. 
 
\subsection{Saddle-node bifurcations of the CP1}\label{sec:SaddleNode}
For the CP1, implementing the constraint \eqref{Gconstr} leads to equation \eqref{eqEquilambdaOriginal}
with coefficients depending on $\E$. Following the arguments of Kummer \cite{Ku} we can look 
for conditions that allow for more than two solutions (in $\lambda$) as an effect of changing $\E$: 
when the number of solutions changes, we have the corresponding {\it bifurcation value} of $\E$. 
The numerator of ${\cal G} (\lambda,\E) = 0$ corresponds to an equation of 4th degree in $\lambda$, namely
\begin{equation}\label{eqEquilambda}
4 (A-\lambda )^2 (B-\lambda )^2\E^2-\delta_1^2  (B-\lambda )^2 -\delta_3^2 (A-\lambda )^2 =0.
\end{equation}

The discriminant  of \eqref{eqEquilambda} with respect to $\lambda$ is a function of $\E$, whose simple zeros provide the required change in the number of solutions. It reads
\begin{equation}
 64 (A - B)^2 \delta_1^2\delta_3^2 \E^2(-108 (A - B)^2 \delta_1^2 \delta_3^2 \E^2 - (\delta_1^2 + \delta_3^2 - 4 (A - B)^2 \E^2)^3) . \label{discriminant}
\end{equation}
For $A\neq B$ and $\delta_1 \delta_3\neq0$, by solving \eqref{discriminant} with respect
to $\E^2$ and disregarding non-real solutions, the bifurcation values of $\E$ can be found by imposing
\begin{equation}\label{eqBifCP1}
f_1(\E) = 0
\end{equation}
where
\begin{equation}
f_1(\E)= 4(A-B)^2\E^2-\left(\delta_1^{2/3} +\delta_3^{2/3}\right)^3.
\end{equation}

For $\Delta_1=0$, we can find two explicit solutions for \eqref{eqBifCP1}, namely
\begin{equation}\label{solES}
\mathcal E_{CP1,\pm}=\frac{\Delta_3}{- D_3\pm \mathcal T_0^3},
\end{equation}
where 
$$ \mathcal T_0=\sqrt{\left(2(A-B)\right)^{2/3}- D_1^{2/3}}\,, $$
provided that $D_3\neq\mp\mathcal{T}_0^3$ and $\mathcal{T}_0^3$ is well defined. Notice that for $D_1=0$ we recover \eqref{enec-inclined-2:2}.

If $\Delta_1\neq0$, we are not able to find explicit analytic solutions of equation \eqref{eqBifCP1} for generic values of the parameters. However, once the parameters are fixed, it is possible to find numerical solutions.

\begin{figure}[htb!]
\begin{center}
\includegraphics[scale=.4]{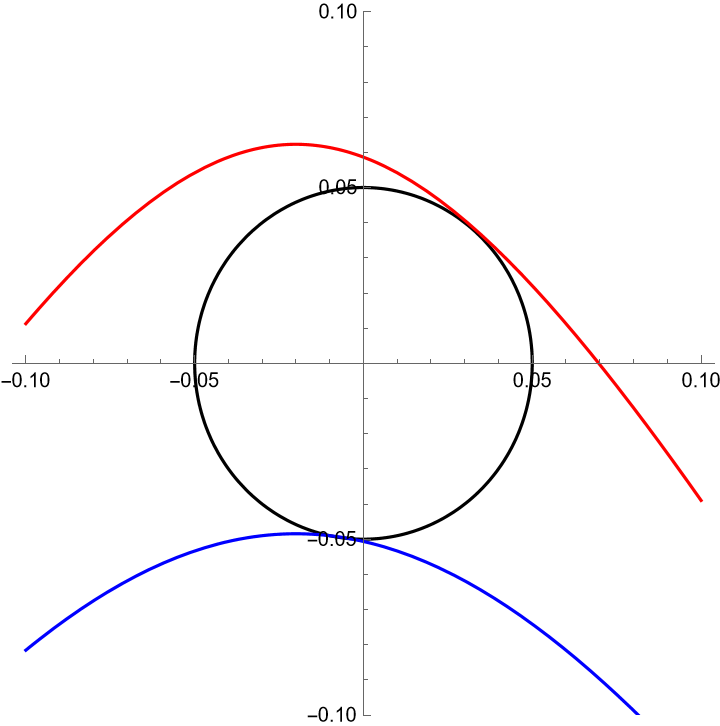}
\qquad
\includegraphics[scale=.4]{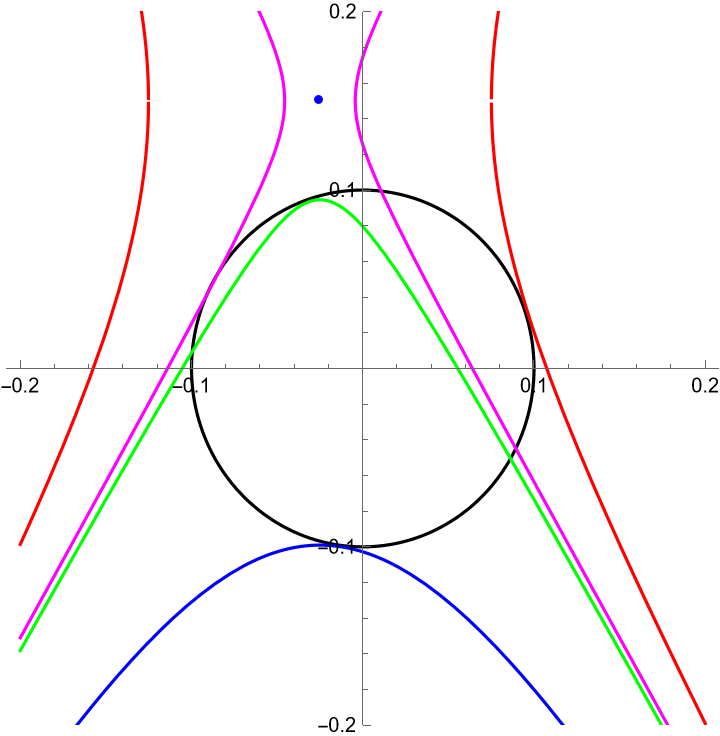}
\qquad
\includegraphics[scale=.4]{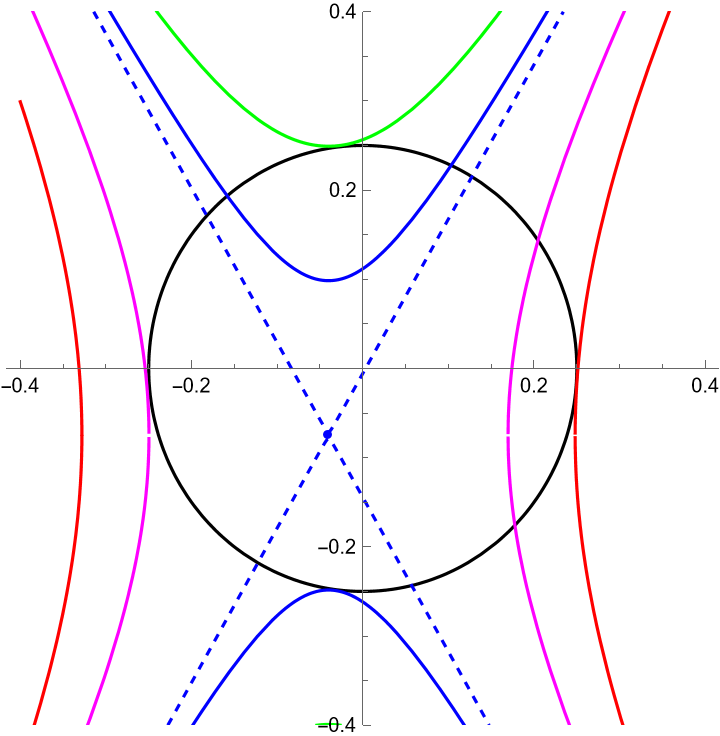}
\qquad
\includegraphics[scale=.4]{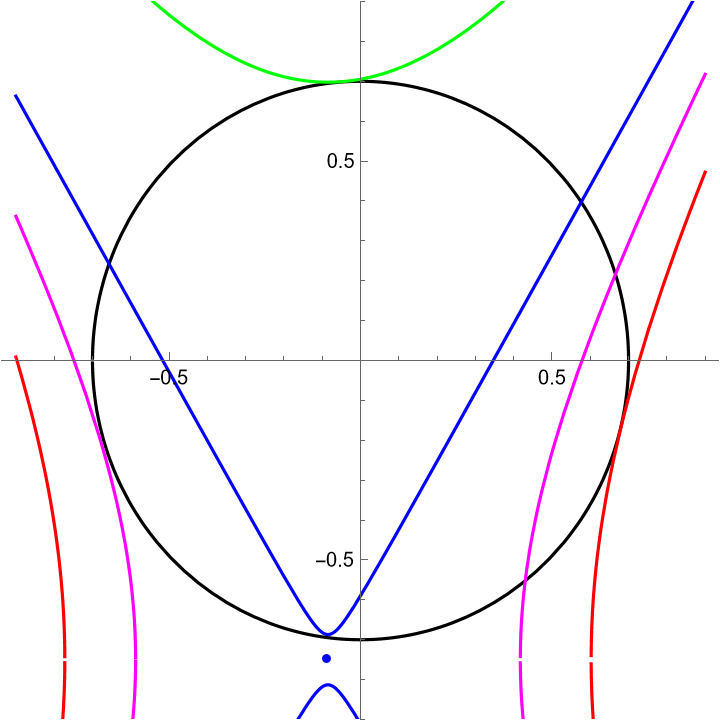}
\end{center}
\caption{Section of the sphere \eqref{rel_invariants} (in black) and the level sets  \eqref{conical} at several values of $h$ on the plane $(I_1, I_3)$ for $A= 1$, $B =-0.\bar 3$, $D_3=-1$, $D_1=0.2$, $\Delta_3=0.2$, $\Delta_1=0.03$ and increasing values of $\mathcal E$. By our choice of the coefficients in the normal form we have $\mathcal E_{CP2,+}\approx 0.12$, $\mathcal E_{CP2,-}\approx 0.59$ and numerically we find only one real positive solution for \eqref{eqBifCP1}, namely $\mathcal E_{CP1,+}\approx0.08$. In the upper left $\E=0.05 <\E_{CP1,+}$: there can be only two external tangency points (see the blue and red hyperbolas) corresponding to two stable equilibria  on the sphere. In the upper right picture  $\mathcal E=0.1>\mathcal E_{CP1,+}$: two new tangency points have appeared, one stable (in magenta) and one unstable (in green); there has been a saddle node bifurcation. At $\mathcal E=\mathcal E_{CP2,+}$ the unstable equilibrium coincides with the center of the hyperbolas that is on the circle now. In the lower left frame,  $\mathcal E=0.25>\mathcal E_{CP2,+}:$ the center (the blue dot) has moved inside the circle. On the sphere it correspond to two unstable equilibria. The dashed blue lines represents the projection of their stable-unstable manifold on the plane $I_2=0$. Furthermore, one of the tangency points on the circle  (with the hyperbola in green) has passed from unstable to stable. Therefore there has been a pitchfork bifurcation. In the lower right picture $\E=0.7>\mathcal E_{CP2,-}$: the center of the family of hyperbolas is outside of the circle. The two unstable equilibria inside the circle have disappeared through another pitchfork bifurcation. 
Since there is only one real positive solution of equation \eqref{eqBifCP1}, there are no other bifurcations. 
\label{fig:hyperbolas1}}
\end{figure}

For example, for the coefficients in the normal form chosen as in Figure \ref{fig:hyperbolas1}, by solving equation \eqref{eqBifCP1} numerically, we find that the only real positive solution is given by $0.0837154$.

We aim here at understanding which kind of bifurcations are related to the solutions of \eqref{eqBifCP1}. We are interested in the case IV) in which $D_j\neq 0$ $\vee$ $\Delta_j\neq 0,$ with $j=1,3$. For each value of the energy rather than the one of the global bifurcation, this request corresponds to the condition $\delta_1\neq 0$ $\wedge$ $\delta_3\neq 0$. We are interested in such a case since the cases $\delta_1\delta_3\equiv0$  (Cases I), II), III)) have already been analyzed. Precisely, we saw that in case $\delta_1=\delta_3=0$ all equilibria of the system are at the poles of the sphere for any value of $\E$ while for $D_1=\Delta_1=0$ ($\delta_1\equiv0$) (resp. $D_3=\Delta_3=0$, i.e. $\delta_3\equiv0$) we know that pairs of equilibria can appear/disappear on the plane $I_2=0$ from the  north/south pole of the sphere (resp. east/west pole) via pitchfork bifurcations (see~\cite{bif11} and Section \ref{sec:bifsymmetric}).

For $\delta_1\delta_3\neq 0$ we know that the CP1 are given by Eq.~\eqref{eqCP1} and that the sphere constraint has equation~\eqref{eqEquilambdaOriginal}, that is well defined.
 As its numerator, given by Eq.~\eqref{eqEquilambda}, is a polynomial equations of fourth degree, the number of its real solutions in $\lambda$ is always even. Since two distinct solutions always exist, i.e. the (QN)-modes (see~\cite{Ku}, \cite{RC23}), we can have a bifurcation only when the number of equilibria changes from two to four, or vice-versa. This can happen at the critical values for $\E$ such that \eqref{eqEquilambda} has a triply and a single root (pitchfork bifurcation) or when  \eqref{eqEquilambda} has two single roots and one double root (saddle-node bifurcation), or when it has two double roots. Let us consider the following possibilities:
 
\begin{equation}\label{eqPitchCP1}
4 (A-\lambda )^2 (B-\lambda )^2\E^2-\delta_1^2  (B-\lambda )^2 -\delta_3^2 (A-\lambda )^2= 4 (\lambda- a)^3(\lambda -b)\E^2
\end{equation}
or
\begin{equation}\label{eqSaddleCP1}
4 (A-\lambda )^2 (B-\lambda )^2\E^2-\delta_1^2  (B-\lambda )^2 -\delta_3^2 (A-\lambda )^2= 4 (\lambda- a)(\lambda -b) (\lambda- c)^2\E^2.
\end{equation}
In the first case two equilibria appear from an existing equilibrium (corresponding to $\lambda=a$). As we shall see in the following, this is possible only in the cases II)- III) and it is associated to  a pitchfork bifurcation \cite{bif11}. In the second case two equilibria appear/disappear at $\lambda=c$. Since at the bifurcation we have two coinciding equilibria and the Euler characteristic of the sphere is two,  by the Poincar\'e-Hopf theorem  \cite{PoiHopfBook}, the two equilibria can not have the same (in)stability. Thus, they appear/disappear via a saddle-node bifurcation. Notice that for $c=a$ or $c=b$ we have triple roots, and therefore we are in the case of a pitchfork bifurcation. In principle, for  $a=b\neq c$ two pairs of equilibria might appear/disappear simultaneously (two double roots). However, as we will see, there are no real solutions for $\E$ that would allow any of these. At the same time, the case $a=b=c$ (quadrupole roots) is not possible, as the system \eqref{hamiltonian_inv_noC} has always at least two distinct equilibria on the sphere \eqref{rel_invariants}, for any $\E>0$ (see~\cite{Ku}).

Let us start from Eq.~\eqref{eqPitchCP1}.
By collecting and equalizing the coefficients of the same degree in $\lambda$ in both left and right hand side of equation \eqref{eqPitchCP1}, we obtain a system of four equations whose solutions provide $a,b$ and the desired critical values of $\E$:
\begin{equation}
\begin{cases}
&-4 a^3 b\E^2-B^2 \left(\delta_1^2-4 A^2 \E^2\right)-A^2 \delta_3^2 =0 \\
&2 a^2 (a+3 b)\E^2 +B \left(\delta_1^2-4 A^2 \E^2\right)-4 A B^2\E^2+A \delta_3^2=0\\
&4  \left(-3 a^2-3 a b+A^2+4 A B+B^2\right)\E^2-\delta_1^2-\delta_3^2=0 \\
&(3 a-2 A+b-2 B)\E^2=0
\end{cases}\, .
\end{equation}
By solving for $a,b,\delta_1, \delta_3$ we find
\begin{equation}\label{eqECP1pitch1}
a=A, \;\; b=2B-A,\;\;\delta_1=0,\;\;\delta_3^2=4(A-B)^2\E^2
\end{equation}
or
\begin{equation}\label{eqECP1pitch2}
 a=B,\;\; b=2A-B,\;\;\delta_3=0,\;\;\delta_1^2=4(A-B)^2\E^2.
\end{equation}
These correspond to the pitchfork bifurcations in the symmetric cases II)-III), though they can not be accepted as solutions for \eqref{eqEquilambdaOriginal}, since we are studying the case IV) when $\delta_1\delta_3\not\equiv 0$ and Eq.~\eqref{eqPitchCP1} is well defined just in this case. However, this proves that the critical points of the first kind cannot appear/disappear through pitchfork bifurcation in the case IV). 

By a similar approach, we can look for a solution of equation \eqref{eqSaddleCP1}. By solving for $a,b,c$ and $\delta_3$, we arrive at the solutions
\begin{equation}\label{SolSaddle-no}
c=B,\;\;\delta_3=0,\;\; b=A\mp\frac{\delta_1}{2 \E},\;\ a=A \pm \frac{\delta_1}{2 \E}
\end{equation}
or
\begin{equation}
\label{SolSaddle}
\begin{split}
a&=A+B-c\pm\frac{\sqrt{8 \E^2 (A-c) (c-B)+\delta_1^2+\delta_3^2}}{2 \E},\\
 b&=-a+2 A+2 B-2 c,\;\; c= A-\frac{\sqrt[3]{\delta_1^2 (A-B)}}{2^{2/3} \E^{2/3}},
\end{split}
\end{equation}
where $\delta_3$ has to satisfy equation \eqref{eqBifCP1}.
Solution \eqref{SolSaddle-no} cannot be accepted as a solution of \eqref{eqEquilambdaOriginal},  because it would require $\delta_3=0$, while we are in the case $\delta_1\delta_3\neq 0$, under which Eq.~\eqref{eqSaddleCP1} is well defined. 

In case $a=b\neq c$, we find
\begin{eqnarray}\label{double}
a=A, \;\; c=B\;\;\hbox{ or }\;\; a=B,\;\;c=A,
\end{eqnarray}
and $\delta_1=\delta_3=0$, not acceptable here. 

Therefore, we conclude that for $A\neq B$, and $\delta_1\delta_3\neq 0$, at the positive critical values of $\E$ that solve equation \eqref{eqBifCP1} we have a saddle-node bifurcation.

Thus,  in  Case IV) a pair of CP1  can appear/disappear only via saddle-node bifurcations, at the critical values of $\E$ that satisfy equation \eqref{eqBifCP1}. As an example, Figures \ref{fig:hyperbolas1}-\ref{fig:hyperbolas2} show a bifurcation sequence in the case $AB<0$ (when the projection of family \eqref{conical} on the plane $I_2=0$ consists of hyperbolas), while Figures \ref{fig:nosymmgb}-\ref{fig:nosymmgbps} refer to the case $AB>0$ (i.e. a case with ellipses).

Notice that for $\delta_1 \delta_3=0$ the discriminant \eqref{discriminant} is identically zero, for each values of $\E$. This confirms that for each values of $\E$ such that $\delta_1 \delta_3=0$, there is no change in the number of equilibria. However, there can be a global bifurcation, see 
Figs. \ref{fig:nosymmgb}-\ref{fig:nosymmgbps}. 

\subsection{Bifurcations in the degenerate cases}
\subsubsection{Bifurcations in the degenerate case $A=B$ }
In case $A=B$ equations \eqref{eqMotion} for $I_2=0$ have the solution
\begin{equation}\label{CP1AugB1}
I_3=\frac{\delta_3}{\delta_1} I_1 ,\;\;I_2=0, \;\; I_1= \pm\frac{|\delta_1| \E}{\sqrt{\delta_1^2+\delta_3^2}}, 
\end{equation}
for $\delta_1\neq0$ and
\begin{equation}
I_1= 0, \;\; I_2=0, \;\; I_3=\pm \E
\end{equation}
for $\delta_1=0$. These are the (QN)-modes that, when the system crosses the $\mathbb Z_2\times \mathbb Z_2$ symmetry, correspond to the normal modes of the $2:2$ resonance (see case II) in Sect. \ref{sec:symmetries0}). Instead, if $\delta_3=0$ the solution becomes
$$
I_1 = \pm \E , \;\; I_2=0, \;\;I_3=0\, .
$$
If $D_1\neq 0$ $\wedge$ $D_3\neq 0$ and $\Delta_1/D_1=\Delta_3/D_3<0$, at the critical value
\begin{equation}
\E=\E_{gb,1}=\E_{gb,2}=- \frac{\Delta_1}{D_1}\,,
\end{equation}
 i.e. when $\delta_1=\delta_3=0$,  we have a circle of non-isolated equilibria on the equator $I_2=0$ of the sphere (see Section~\ref{sec:equidegAB}). These are critical points of the first kind ($I_2=0$). Higher order terms in the normal form would be needed to describe the corresponding bifurcation. 

For $I_2\neq0$, we find
the pair of CP2
\begin{equation}\label{CP2AugB}
I_1=-\frac{\delta_1}{2 A},\;\; I_2= \pm\frac{\sqrt{4 A^2 E^2-(\delta_1^2+\delta_3^2)}}{2 A},\;\; I_3=-\frac{\delta_3}{2A}
\end{equation}
corresponding to the center of the family of ellipsoidal cylinders \eqref{conical} that enters/leaves the sphere \eqref{rel_invariants} via pitchfork bifurcations when
\begin{equation}
\label{eq.R}
4 A^2 \E^2-(\delta_1^2+\delta_3^2)=0.
\end{equation}
By solving the above equation, we find two critical values (if $D_1^2+D_3^2\neq 4 A^2$)
\begin{equation}
\label{en.R}
\E=-\frac{D_1\Delta_1+ D_3 \Delta_3 \pm \sqrt{4 A^2 \left(\Delta_1^2+\Delta_3^2\right)- (D_3 \Delta_1-D_1 \Delta_3)^2}}{D_1^2+D_3^2-4A^2}
\end{equation}
that, when well defined and not negative, coincides with \eqref{sol.CP2} for $A=B$.

\subsubsection{Bifurcations in the degenerate case $AB=0$}

Let us start by the case $A=0$, $B\neq0$. 
The level sets \eqref{conical} are parabolas on the plane $I_2=0$ with vertex at 
$I_3=-\frac{\delta_3}{2B}$. For $\delta_1=0$ they break into  two straight lines parallel to the $I_1$ axis. At $h=-\frac{\delta_3^2}{4B}$ the two lines coincide and we have a circle of non-isolated equilibria on the plane $I_2=0$. 
Therefore, to understand this bifurcation, a higher order normal for is needed.
As long as $\delta_1\neq0$ equation \eqref{crit_eq} can be solved only if $I_2=0$, i.e. we can have only CP1. The critical values of $\E$ that determine the related saddle-node bifurcation are given by the solutions of \eqref{eqBifCP1} for $A=0$. 

\noindent
The case $B=0$, $A\neq0$ is similar. The circle of equilibria would appear in this case if $\delta_3=0$.

\noindent
In case $A=B=0$ the normal form becomes a linear function, therefore the level sets \eqref{conical} degenerate into planes. Thus, we can have only two tangency points on the circle \eqref{circumference}, namely at
\begin{equation}\label{normalmodes0}
I_3=\pm \frac{\delta_3 \E}{\sqrt{\delta_1^2+ \delta_3^2}}\, ,
\end{equation}
corresponding to the two stable (QN)-modes, that  for $\delta_1=0$ or $\delta_3=0$ are normal modes.

\begin{figure}[htbp]
\includegraphics[scale=.4]{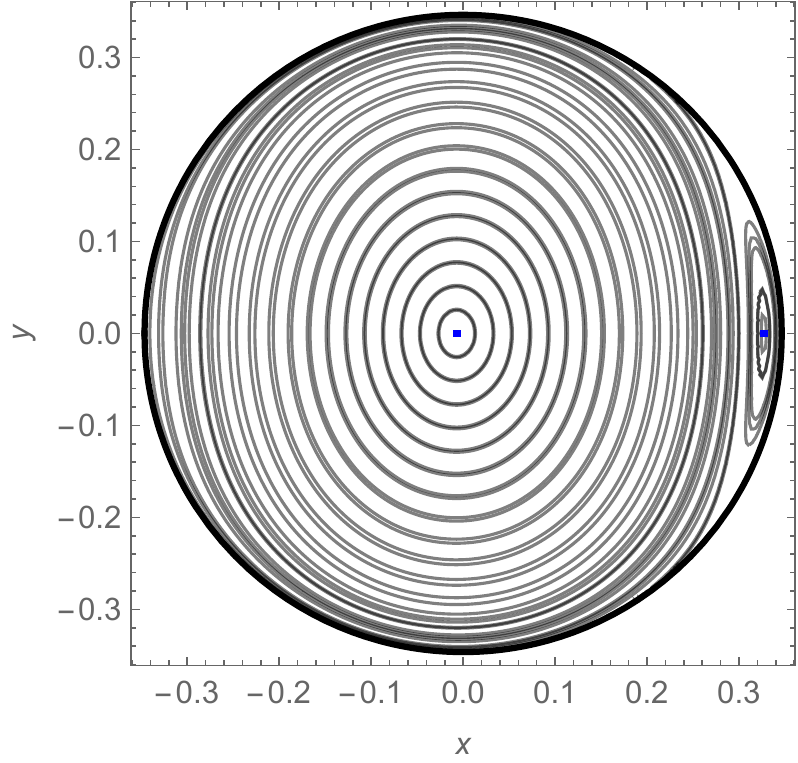}
\quad
\includegraphics[scale=.4]{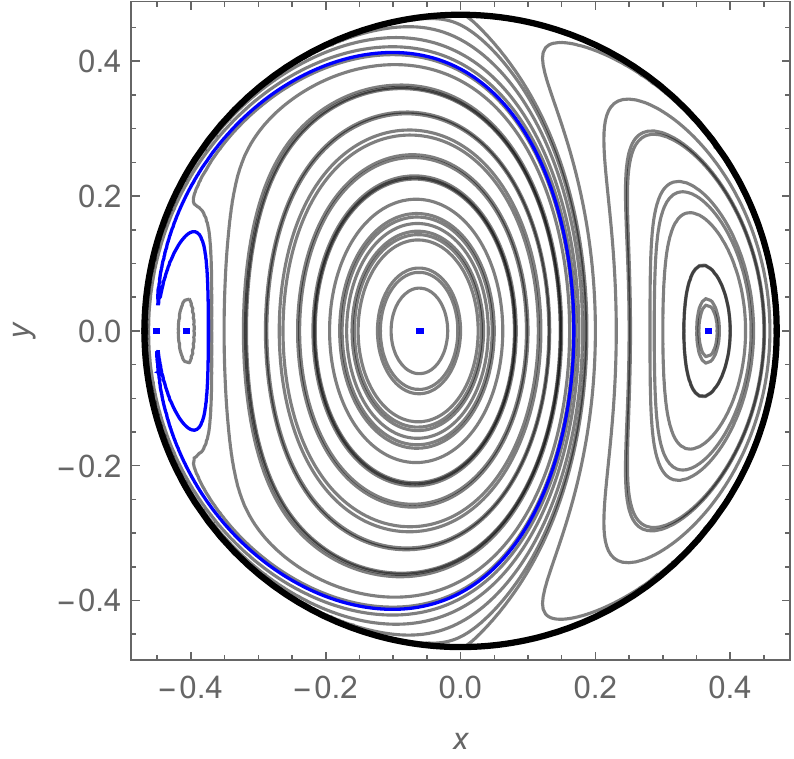}
\quad
\includegraphics[scale=.4]{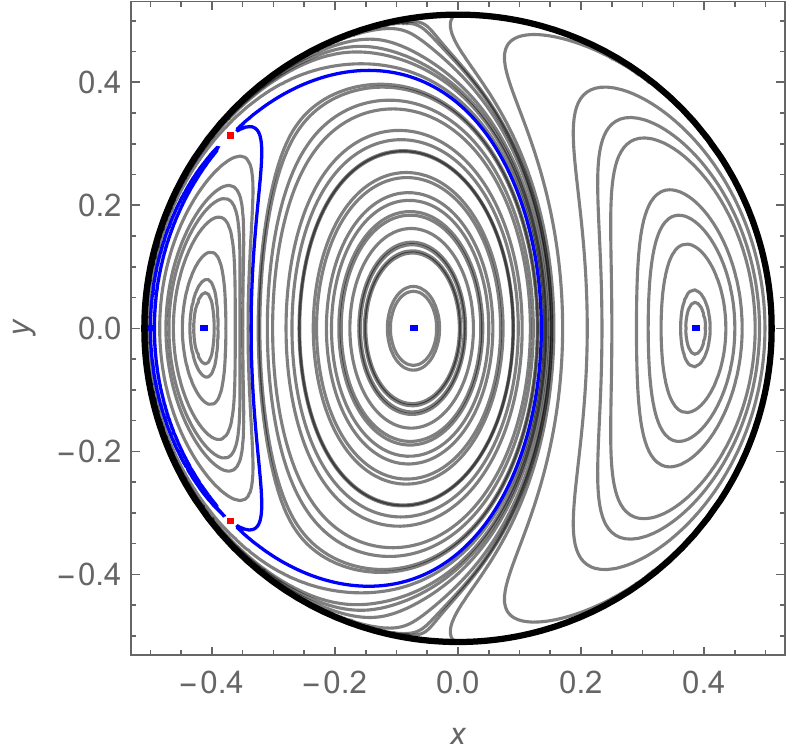}
\\
\includegraphics[scale=.4]{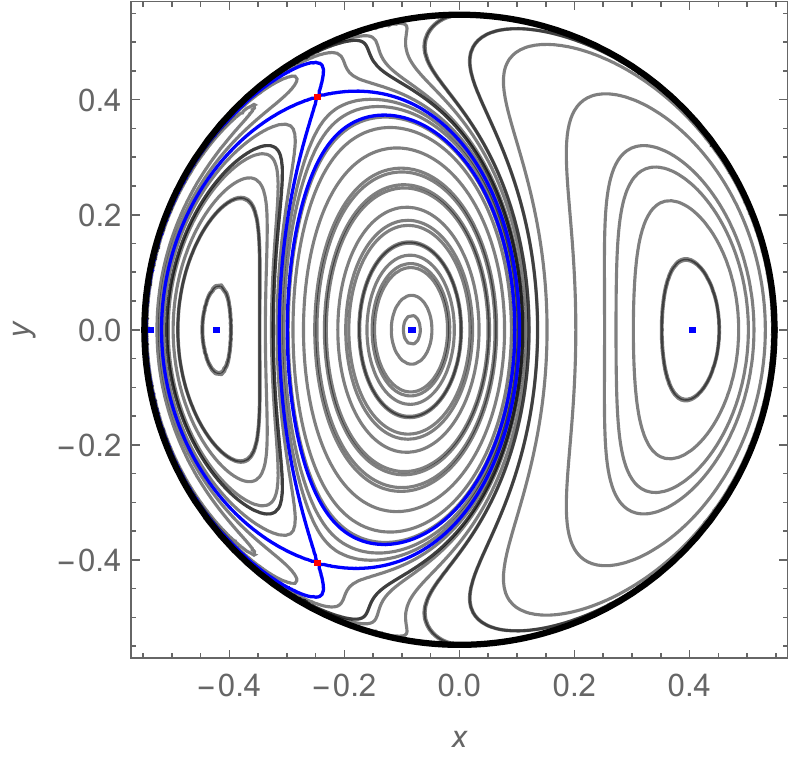}
\quad
\includegraphics[scale=.4]{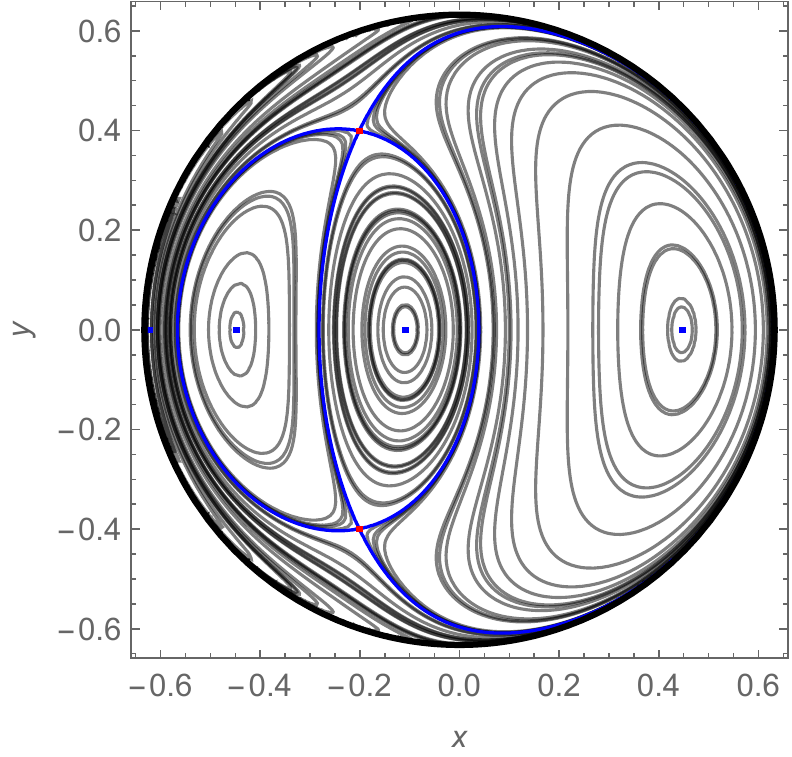}
\quad
\includegraphics[scale=.4]{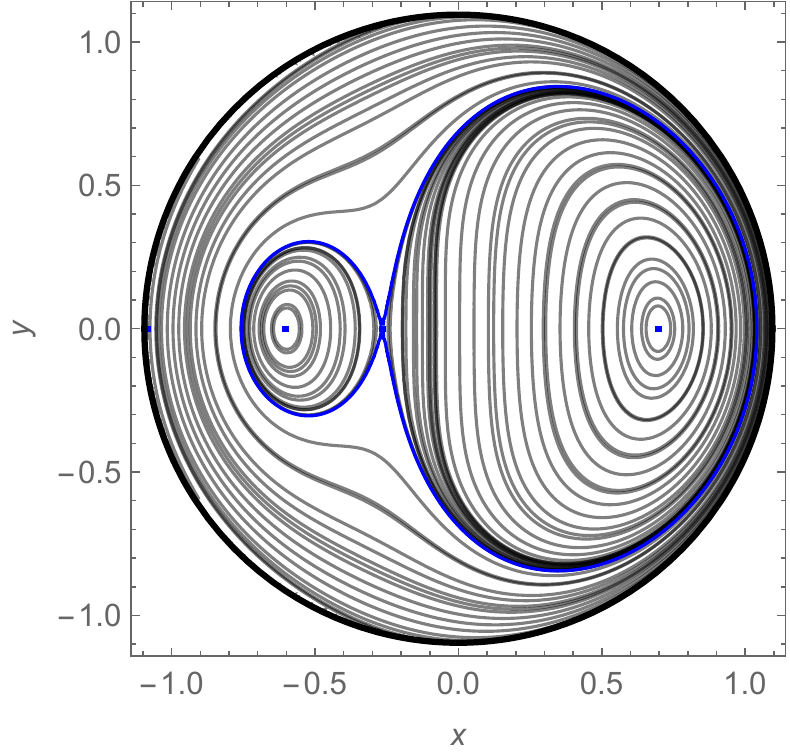}
\caption{Surfaces of section in the $(x,y)$ stereographic coordinates defined in \eqref{ccoord} corresponding to the cases of Fig.~\ref{fig:hyperbolas1}. The values of the distinguished parameter are: $ \E=0.06$ (top left panel), $ \E=0.11$ (top central panel), $ \E=0.13$ (top right panel), $ \E=0.15$ (bottom left panel), $ \E=0.2$ (bottom central panel), $ \E=0.6$ (bottom right  panel). To make more visible the saddle-node bifurcation, we chose here $D_1=0.9$ instead of $0.2$ as Fig.~\ref{fig:hyperbolas1}. The bifurcation sequence does not change anyway.}
\label{fig:hyperbolas2}
\end{figure}

\section{Catastrophe map and bifurcation sequences}\label{sec:catamap}
In analogy with what done in \cite{bif11}, for $2B\neq A$ we define the parameters:
\ba
\alpha &=& \frac{A}{2B - A}, \label{CPA} \\
\beta_1 &=& \frac{\delta_1}{(2B - A)\E} = \frac{D_1\E +\Delta_1}{(2B - A)\E}, \label{APA1} \\
\beta_3 &=& \frac{\delta_3}{(2B - A)\E} = \frac{D_3\E +\Delta_3}{(2B - A)\E}. \label{APA3}
\ea
We can then think to represent the bifurcation sequences of the various families in the 3-dimensional space $(\alpha,\beta_1,\beta_3)$, that we call 3DPS. The parameter $\alpha$ is the \sl coupling parameter, \rm whereas $\beta_1,\beta_3$ are the \sl asymmetry parameters \rm (we could even say \sl detuning \rm parameters). 

\noindent
Once specified a model given by $A,B,D_1,\Delta_1,D_3,\Delta_3$, in the 3DPS a straight line parameterized by $\E$ ensues, perpendicular to the $\alpha$-axis. When this line intercepts a \sl bifurcation locus \rm a qualitative change in the phase-space structure is produced. 

Before examining the general case, let us see how the bifurcation loci appear on the two coordinate planes 
$(\alpha,\beta_3)$ and $(\alpha,\beta_1)$.

\subsection{$\delta_1\equiv0$}\label{sec:catmapdelta10}

As said in Sect. \ref{sec:symmetries0}, the case $\delta_1\equiv0$, $\delta_3$ not identically zero, is the same as in \cite{bif11}. 

\begin{figure}[htp]
\begin{center}
\includegraphics[scale=.5]{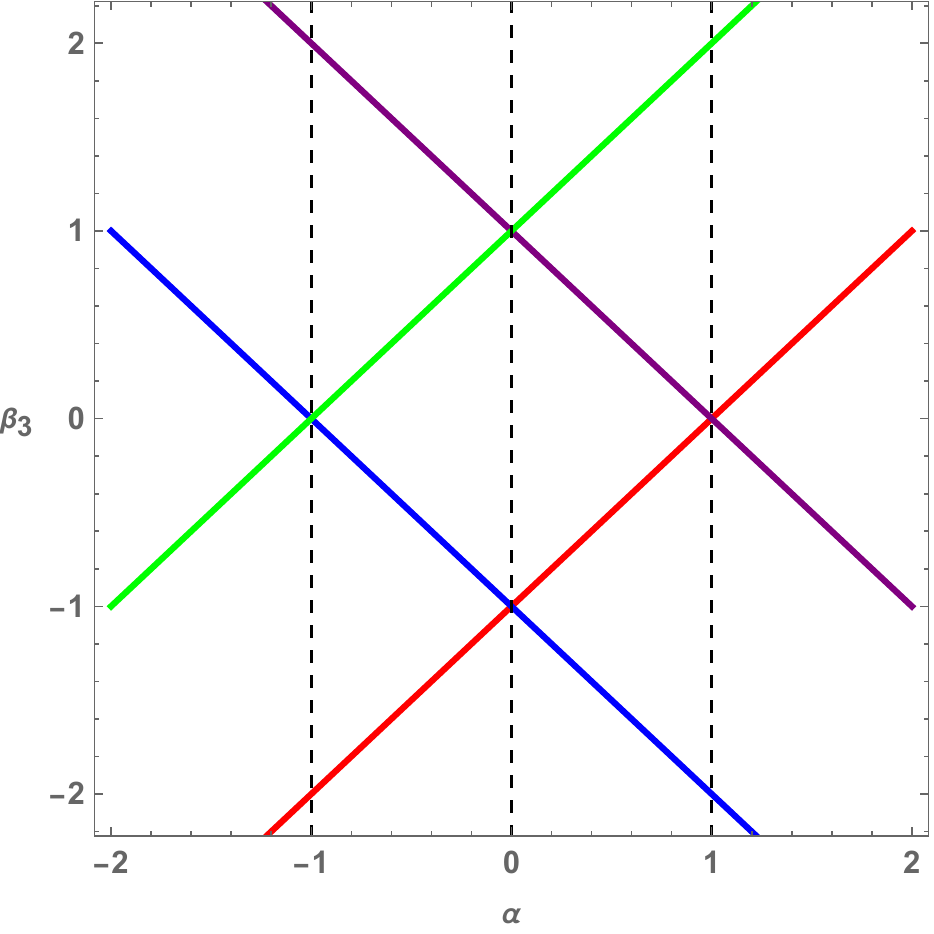}
\includegraphics[scale=.51]{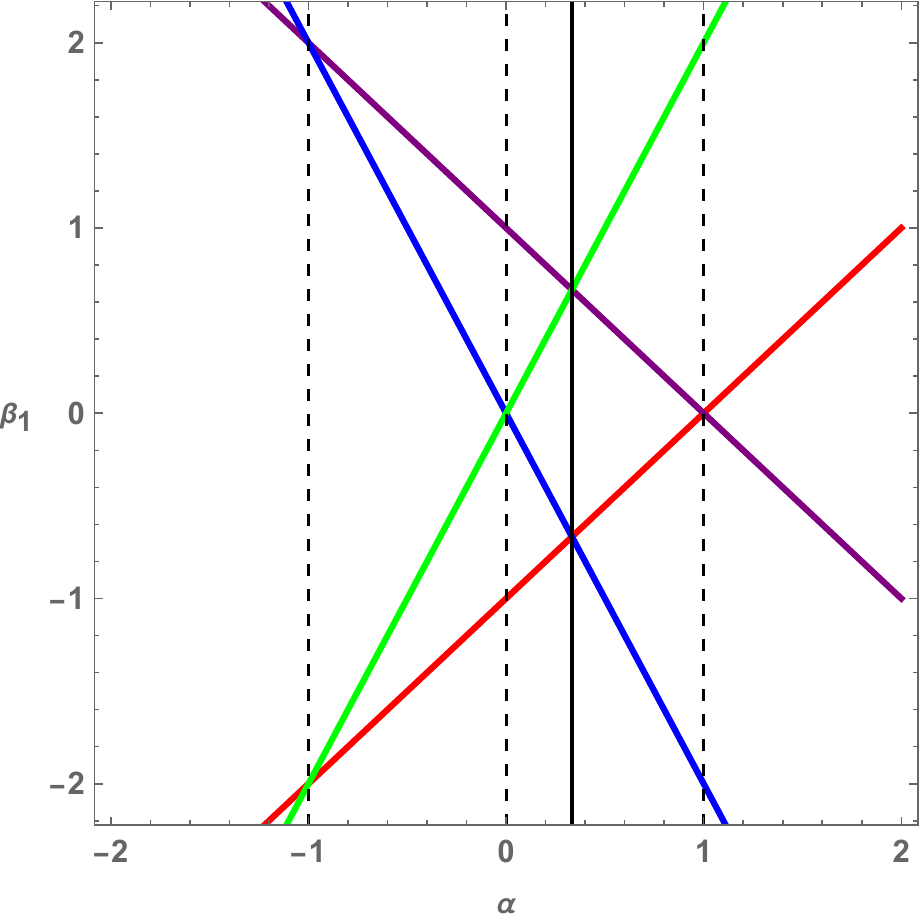}
\end{center}
\caption{Catastrophe map, on the left the case $\delta_1\equiv0$, on the right the case $\delta_3\equiv0$. The red and purple lines correspond to the pitchfork bifurcations of the CP1, while the blue and green lines determine the pitchfork bifucation of the CP2. They intersects at the degenerate cases $\alpha=-1,0,1$. If $\delta_3=0$, we see another intersection at $\alpha=\frac13$ (the black line in the right figure). The black line at $\alpha=\frac13$ marks a a structurally unstable situation in which  a pair of CP1 and a pair of CP2 appear simultaneously at the intersection points of the bifurcation lines. See Figures \ref{fig:BifCurves2D} (right) and \ref{fig:CatMapAlpha01} (middle) for the analogous scenario in the generic case $\delta_1\delta_3{\not\equiv}0$.}\label{CmapB3}
\end{figure}

The bifurcation thresholds \eqref{enec-inclined-2:2} and \eqref{enec-loop-2:2} become, when expressed on the 
$(\alpha,\beta_3)$-plane, the two pairs of straight lines
\ba
\beta_3 (\E_{ i, 1\pm})&=& \pm (1 - \alpha), \label{B3U} \\
\beta_3 (\E_{\ell, 1\pm}) &=& \pm (1 + \alpha). \label{B3L}
\ea
We obtain the \sl catastrophe map \rm (the term comes from the `chemical literature', see e.g. \cite{STK}) in the left panel of Fig.\ref{CmapB3}: red and purple lines correspond to \eqref{B3U}; green and blue lines to \eqref{B3L}.

Each vertical line represents a system depending on $\E$. The intersections with the bifurcation lines correspond to the birth and/or annihilation of a pair (due to the symmetry) of periodic orbits in generic position via pitchfork bifurcations. The dashed lines at $\alpha=-1,0,1$ mark the degenerate cases at $B=0$, $A=0$ and $A=B$, respectively (cf. Sections \ref{sec:equidegAB}-\ref{sec:equidegABzero}).

\subsection{$\delta_3\equiv0$}\label{sec:catmapdelta3zero}
The case III) ($\delta_3\equiv0, \delta_1 \not\equiv0$) is analogous but with non-trivial differences. 

The bifurcation sequences \eqref{E1scambio} and \eqref{E2scambio} become, when expressed on the 
$(\alpha,\beta_3)$-plane, the two pairs of straight lines
\ba
\beta_1 (\E_{ i, 2\pm})&=& \pm (1 - \alpha), \label{B1U} \\
\beta_1 (\E_{\ell, 2\pm}) &=& \mp 2 \alpha. \label{B1L}
\ea
We then obtain the catastrophe map in the right panel of Fig.\ref{CmapB3}: red and purple lines correspond to \eqref{B1U}, while green and blue lines to \eqref{B1L}. As before, each vertical line represents a system depending on $\E$. The intersections with the bifurcation lines correspond to the birth and/or annihilation of a pair of periodic orbits. The continuous black line corresponds to $\alpha=1/3$ which is given by $B=2A$. In this case, at the intersection point of the bifurcation lines a pair of CP1 and a pair of CP2 appear simultaneously. This is a structurally unstable situation in which higher order terms in the normal form \eqref{hamiltonian_inv_noC} would be needed to discern the bifurcation scenario.

\subsection{$\delta_1\delta_3\not\equiv0$}\label{sec:catmapgeneric}
\begin{figure}[htpb]
\begin{center}
\includegraphics[scale=.5]{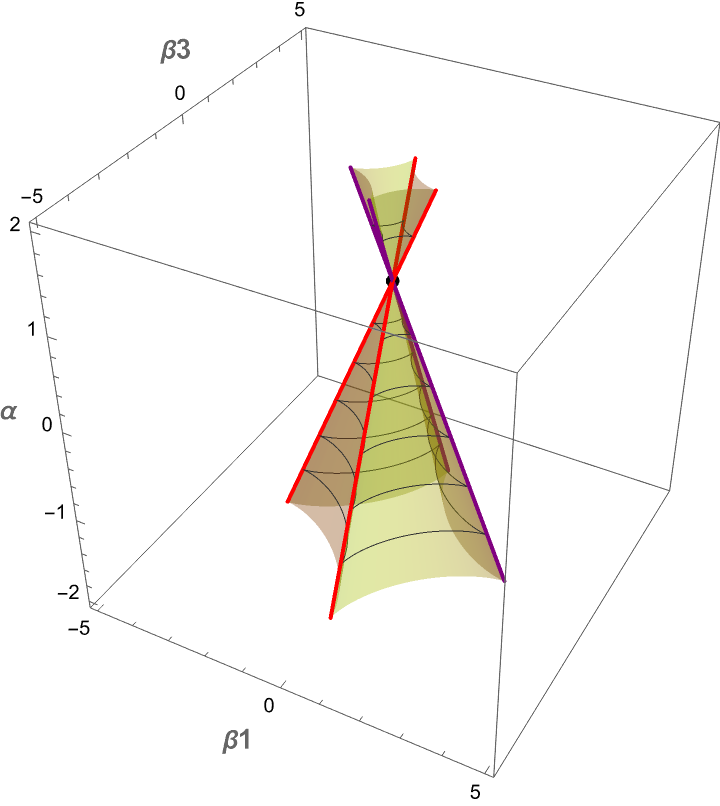}
\qquad
\includegraphics[scale=.5]{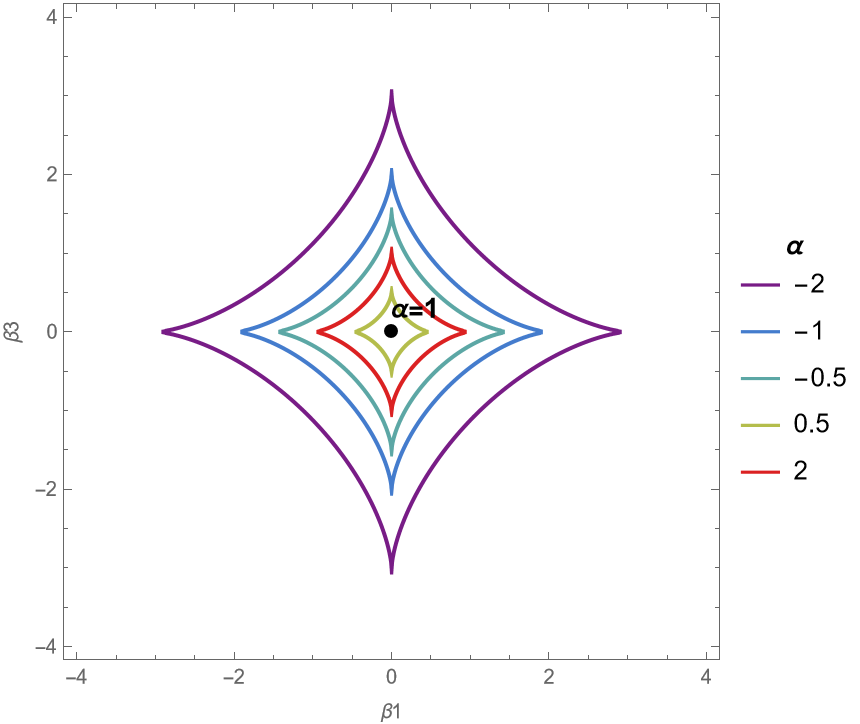}
\end{center}
\caption{The surface that determines the bifurcation of the CP1. Left: plot in the 3DPS. The surface corresponds to saddle- node bifurcations, except at the red and purple lines that mark the pitchfork bifurcations at $\delta_1 \delta_3\equiv0$, 
see also Figure \ref{CmapB3}. They intersect at the singularity (black point) at $\alpha=1$. Right: contour plot on the $(\beta_1,\beta_3)$ plane. } \label{fig:BifCurvesCP1}
\end{figure}

The bifurcation surfaces \eqref{eqBifCP2}, \eqref{eqBifCP1}, expressed in the $(\alpha,\beta_1,\beta_3)$ parameters, read
\begin{equation}\label{eqBifCP2AlphaBeta}
\frac{\beta_1^2}{4\alpha^2}+ \frac{\beta_3^2}{(1+\alpha)^2}=1
\end{equation}
and
\begin{equation}\label{eqBifCP1AlphaBeta}
\beta_1^{\frac23}+\beta_3^{\frac23}=(1-\alpha)^{\frac23}\,,
\end{equation}
respectively (see Figure \ref{fig:BifCurvesCP1} for  the surface corresponding to \eqref{eqBifCP1AlphaBeta}). The surface is the union of two ``star-shaped'' cones that meet at at their tip point. At the regular points of the cones, we have saddle-node bifurcations for the CP1. Referring to Fig. \ref{fig:BifCurvesCP1}, at the red and purple lines $\beta_1\beta_3=0$, thus we have pitchfork bifurcations (see also Figure \ref{CmapB3}). The lines of pitchfork bifurcations meet at the tip point where  we are at the degenerate case $\alpha=1$ (when the CP1 degenerate into a circle of non-isolated equilibria).  

\begin{figure}[htpb]
\begin{center}
\includegraphics[scale=.5]{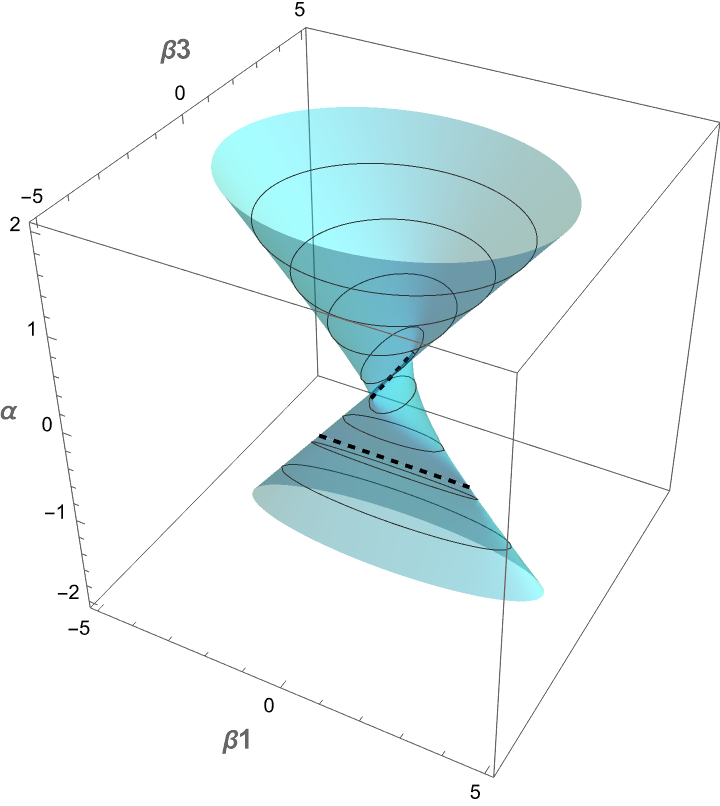}
\qquad
\includegraphics[scale=.5]{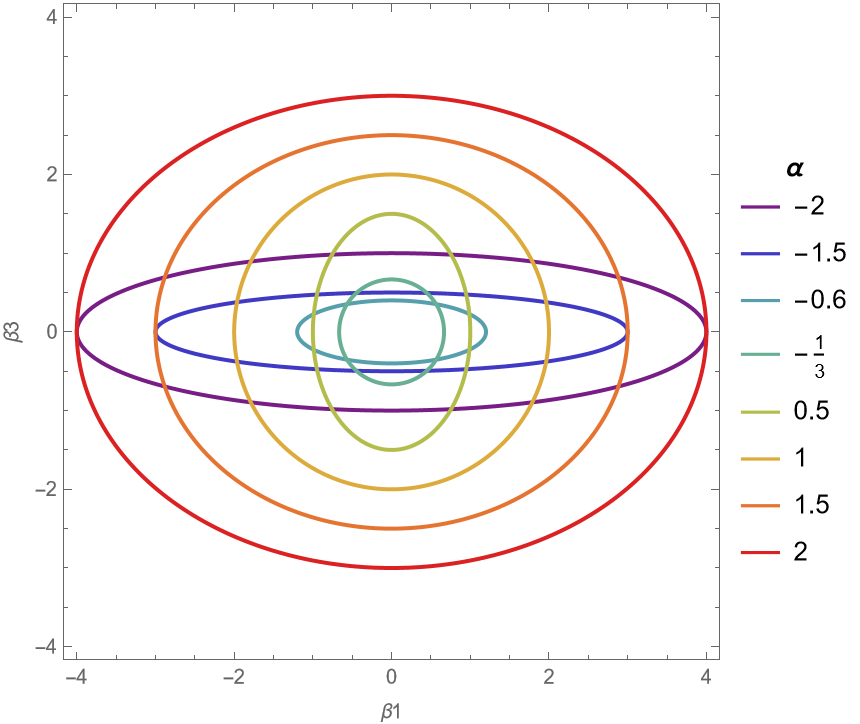}
\end{center}
\caption{The surface that determines the appearance/disappearance of the CP2 via pitchfork bifurcations. Left: plot in the 3DPS. The dashed lines correspond to the singularities at $\alpha=0, \beta_1=0$ and at $\alpha=-1, \beta_3=0$, where the surface \eqref{eqBifCP2AlphaBeta} contains only the points at $\beta_3=\pm1$  and at $\beta_1=\pm2$, respectively (where the CP2 degenearate to a circle of non-isolated equilibria). Right: contour plot on the $(\beta_1,\beta_3)$ plane. For $\alpha<-\frac13$ and $\alpha>1$ the ellipses are oblate. We have circles at $\alpha=\frac13$ and $\alpha=1$ and prolate ellipses when $-\frac13<\alpha<1$.} \label{fig:BifCurvesCP2}
\end{figure}

The surface $\eqref{eqBifCP2AlphaBeta}$ of pitchfork bifurcations of the CP2 is represented in Figure \ref{fig:BifCurvesCP2}. We have the union of cylinder like surfaces  that throttle when approaching the lines $\alpha=0$ and $\alpha=-1$ where the CP2 degenerate into a circle of non-isolated equilibria. 

\begin{figure}[htpb]
\begin{center}
\includegraphics[scale=.5]{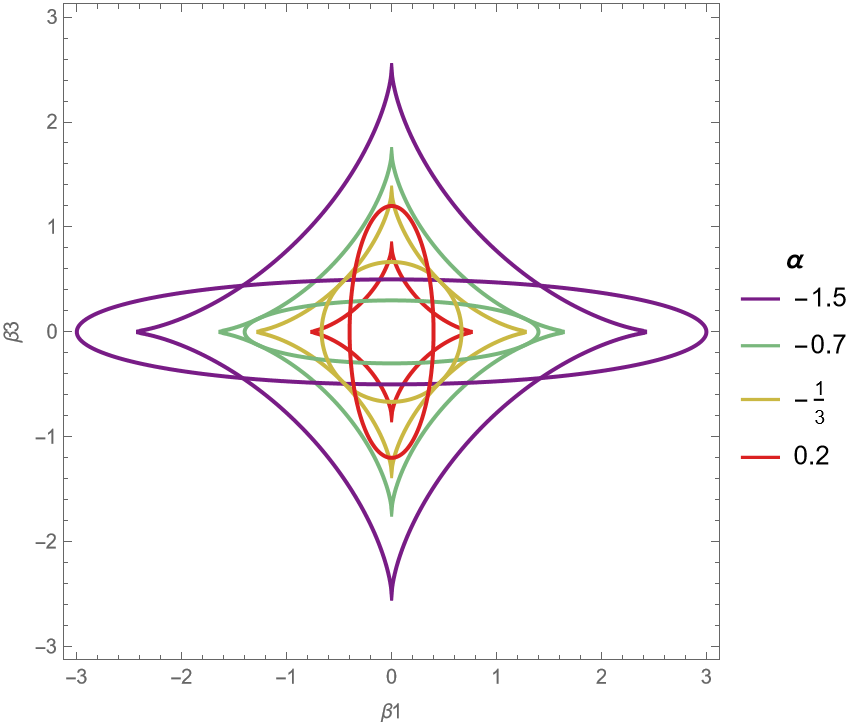}
\qquad
\includegraphics[scale=.5]{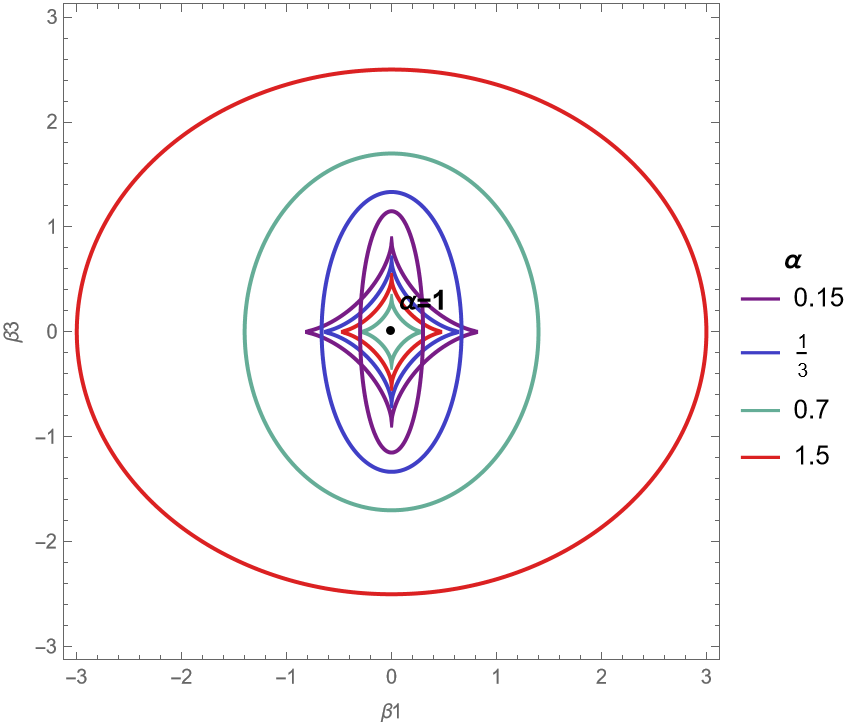}
\end{center}
\caption{Contour plots of both surfaces \eqref{eqBifCP2AlphaBeta}, \eqref{eqBifCP1AlphaBeta} at different fixed values of $\alpha$. } \label{fig:BifCurves2D}
\end{figure}

On the $(\beta_1,\beta_3)$ plane, at fixed $\alpha$, the surface \eqref{eqBifCP2AlphaBeta} corresponds to an ellipse (see Figure \ref{fig:BifCurvesCP2}, right), while the surface  \eqref{eqBifCP1AlphaBeta} is given by a star-shaped figure (see Figure \ref{fig:BifCurvesCP1}, right).
 In Figure \ref{fig:BifCurves2D} we show both surfaces \eqref{eqBifCP1AlphaBeta}, \eqref{eqBifCP2AlphaBeta} together in the $(\beta_1,\beta_3)$, at different fixed values of $\alpha$. Depending on $\alpha$, the two surfaces can intersect at four points, be tangent, or have no contact points. Namely, one can easily find analytically that 
\begin{enumerate}
\item the two curves are tangent, with an in-ellipse, for $-1<\alpha<0$;
\item the two curves have four intersection points (that are not tangency points) if 
\begin{itemize}
\item $\alpha<-1$,
\item $0<\alpha<\frac13$;
\end{itemize}
\item the two curves intersect at the vertices of the star-shaped figure when $\alpha=\frac13$;
\item the two curves do not have contact points for  $\alpha>\frac13$.
\end{enumerate}

At a fixed value of $\alpha$, we can investigate the possible bifurcation sequences on the $(\beta_1,\beta_3)$ plane. On this plane, each system is represented by a (segment of) line parametrized by $\E$. Each line ends at the point ``at infinity'', given by
\begin{equation}
P_{\infty}\equiv\left(\frac{D_1}{2B-A}, \frac{D_3}{2B-A}\right)\,\label{P0}
\end{equation}
and found by taking the limit of \eqref{APA1} and \eqref{APA3} for $\E\to\infty$. 
The slope of the line depends on $\Delta_1$ and $\Delta_3$. Namely, the line is vertical for $\Delta_1=0$, otherwise its slope is given by 
\begin{equation}
m=\frac{\Delta_3}{\Delta_1}\,.\label{angularCoeff}
\end{equation}

In Figure \ref{fig:CatMapFigHyperbola}, the catastrophe map corresponding to the bifurcation sequence of Figures \ref{fig:hyperbolas1}-\ref{fig:hyperbolas2} is represented. In green we have the line corresponding to the system when the parameters are chosen as in Figure \ref{fig:hyperbolas1}. When the line intersects the contour of the surface \eqref{eqBifCP2AlphaBeta} or \eqref{eqBifCP1AlphaBeta} the system passes through a bifurcation. For increasing values of $\E$, we read the sequences moving upward towards the point $P_{\infty}$ (Eq.~\eqref{P0}). Thus, the line crosses at first the curve of the saddle-node bifurcations, then twice the ellipses of the pitchfork bifurcation: a pair of CP1 appears after a saddle-node bifurcation, then we have a pair of CP2 that appears via a pitchfork bifurcation and then disappears via an anti-pitchfork bifurcation. Since in this case $AB<0$, the CP2 are unstable. This implies that the pitchfork bifurcation involved the unstable CP1 that appeared after the saddle-node bifurcation and becomes stable after the appearance of the unstable CP2, as shown in Figures \ref{fig:hyperbolas1}-\ref{fig:hyperbolas2}.

However, depending on $\alpha$, other bifurcation scenarios are possible, according to the mutual position of the curves  \eqref{eqBifCP2AlphaBeta}, \eqref{eqBifCP1AlphaBeta} and the line corresponding to the system in the $(\beta_1, \beta_3)$ plane.
Moreover, the number of intersections between the line and the curves of bifurcation also depends on the slope $m$ of the line (depending on $\Delta_1$ and $\Delta_3$) and on the position of the point $P_{\infty}$ at $\E$ infinity, that for fixed $\alpha$ depends on $D_1$ and $D_3$.
However, up to a rotation on the sphere \eqref{rel_invariants} that exchanges $I_1$ with $-I_1$ and $I_3$ with $-I_3$, we can always assume both $\Delta_1$ and $\Delta_3$ not negative. Furthermore, up to a rotation that exchanges $I_1$ and $I_3$, we can set $\Delta_1\leq\Delta_3$. Therefore, from now on, we assume
\begin{equation}
\Delta_3\geq\Delta_1\geq0 \label{DeltaAssumptions}
\end{equation}
and thus the slope $m\geq1$.

In the followin, we describe the possible bifurcation scenarios depending on $\alpha$, focusing on the bifurcation sequences that are closer to the ones of the $2:2$ resonance (Case II) with the assumption \eqref{DeltaAssumptions}. However, from our approach, any possible bifurcation sequence for the system can be easily reconstructed.  By excluding the degenerate cases ($\alpha=-1,0,1$) we find four intervals for $\alpha$ that determine $4$ major cases, which are investigated in the following subsections.

Since $\alpha$ is defined when $2B\neq A$, for the sake of simplicity, we shall consider only the case in which $D_3(2B-A)>0$, when the point at infinity has a positive $\beta_3$-coordinate and the bifurcation sequences always start when $\beta_3<0$, since $\beta_3\to-\infty$ for $\E\to0$.  The case $D_3(2B-A)>0$ would follow in a similar way.

\subsubsection{$-1<\alpha<0$}
Let us start by considering the case in which the level sets of the normal form \eqref{hamiltonian_inv_noC} are hyperbolas, i.e. the case when  $AB<0$. This is possible only when  $-1<\alpha<0$.
This turns out to be also the only case in which the two curves of bifurcations \eqref{eqBifCP1} and \eqref{eqBifCP2} have tangency points.

For $-1<\alpha<0$, on the plane $(\beta_1,\beta_3)$ the ellipses that determine the pitchfork bifurcations are always internally tangent to the star-shaped curves (see Figure \ref{fig:BifCurves2D}, left). In the limit for $\E\to0$, unless $\Delta_1=\Delta_3=0$, at least one between $\beta_1$ and $\beta_3$ has an infinite value (positive or negative). Therefore, increasing $\E$, the first bifurcation (if any) occurs when the line representing a system crosses the star-shaped curve. 

\begin{figure}[htpb]
\begin{center}
\includegraphics[scale=.5]{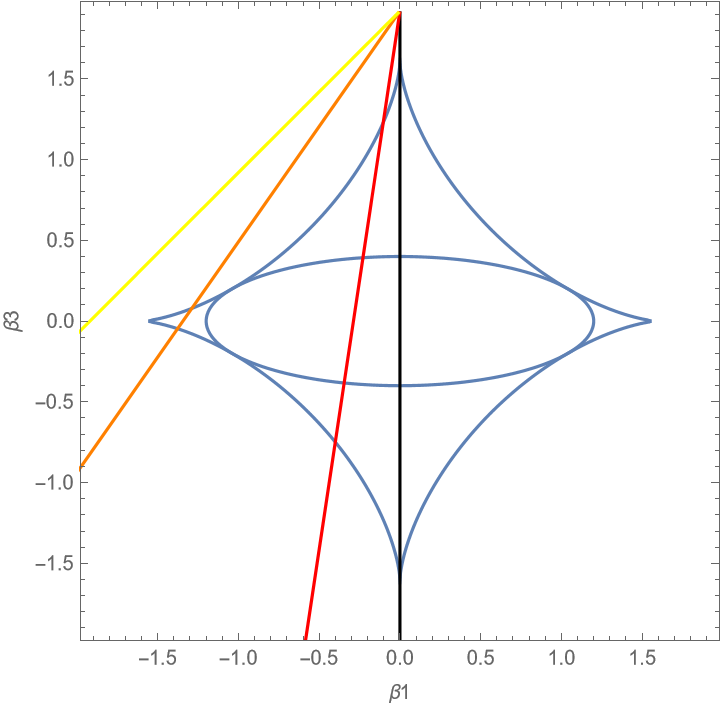}
\quad
\includegraphics[scale=.5]{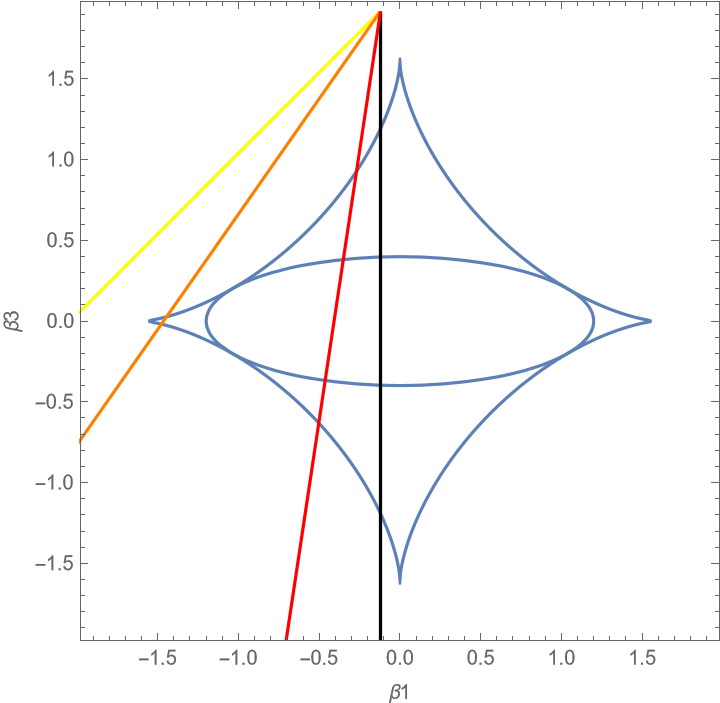}
\quad
\includegraphics[scale=.5]{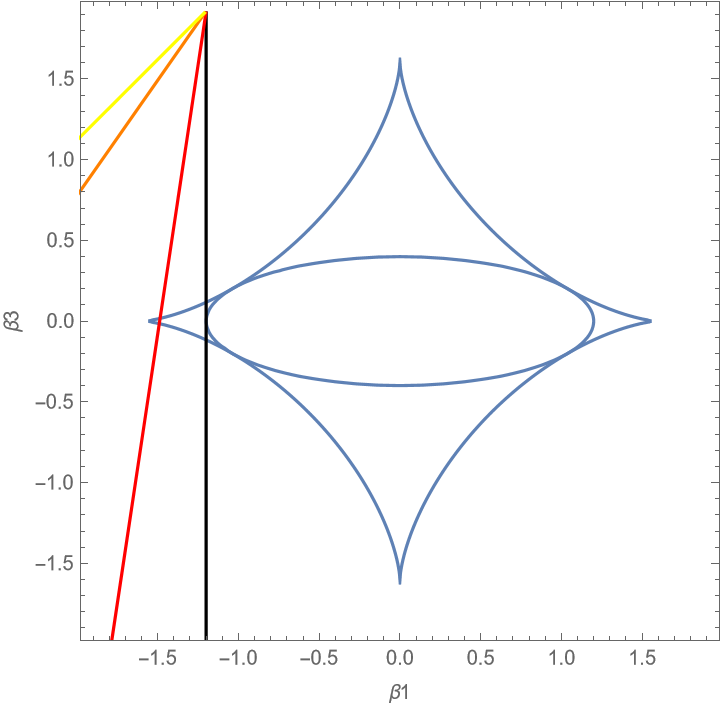}
\includegraphics[scale=.5]{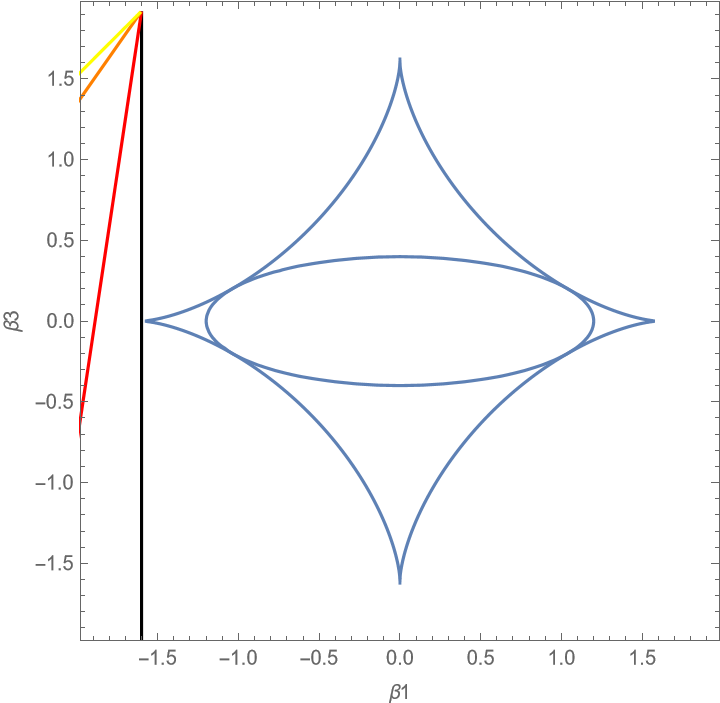}
\end{center}
\caption{Catastrophe maps for $\alpha=-\frac35$ ($A=1, B=-\frac13$), $D_3=-3.2$, $\Delta_3=0.2$. First row: from left to right, $D_1=0$, $D_1=0.2$. Second row: from left to right, $D_1=2$, $D_1=\frac83$. The straight lines represent different systems at different values of $\Delta_1$: $\Delta_1=0$ (black line), $\Delta_1=0.03$ (red line), $\Delta_1=0.14$ (orange line), $\Delta_1=0.2$ (yellow line).} \label{fig:CatMapD1p}
\end{figure}

If $D_1=\Delta_1=0$ the system is represented by the straight line $\beta_1\equiv0$: the  bifurcation sequence follows the one of Case II) (the  $2:2$ resonance case) (see Figure \ref{fig:CatMapD1p}, first frame, black line). The position of the point $P_{\infty}$ (which for fixed $\alpha$ depends only on $D_1$ and $D_3$) determines the end point of the bifurcation sequence.  
Thus, for a parameter choice as in  Figure \ref{fig:CatMapD1p} the bifurcation sequence of the $2:2$ resonance has four steps. 
We know that the so called inclined orbits (that are CP1)  and  the so called loop orbits (that are CP2) appear/disappear from the two normal modes of the system via pitchfork bifurcations \cite{bif11}. Since $-1<\alpha<0$ we have $AB<0$. Thus, the level sets of the Hamiltonian are hyperbolas, with the CP2 always unstable.

In fact, when the black line intersects the star-shaped curve, we have a pitchfork bifurcation from one of the normal modes, followed by another pitchfork bifurcation when the line crosses the ellipse. Since the CP2 are unstable, the two pitchfork bifurcation must involve the same normal mode, that comes back to stability after the appearance of the pair of unstable CP2. The sequence then continues with two anti-pitchfork bifurcations that determine first the disappearance of CP2 and then the disappearance of CP1 from the second normal mode \cite{bif11}.

At fixed $\Delta_3\geq\Delta_1\neq0$, the value of $\Delta_1$  determines the slope of the  line. The smaller $\Delta_1$, the greater is the slope  (see Figure \ref{fig:CatMapD1p} for $D_1=0$, but $\Delta_1\neq0$). Therefore, the greater is $\Delta_1$, the more the line, and thus the bifurcation sequence, is far from the one of the $2:2$ resonance.

Let us see how the bifurcation sequence differs from the one of the $2:2$ resonance as $\delta_1\neq0$. By increasing the value of $\E$, let us read the  bifurcation sequence corresponding to the red line, that is very close to the one of the $2:2$ resonance ($D_1=0$, $\Delta_1=0.03$, Figure \ref{fig:CatMapD1p}, top left).  At values of the energy $\E$ small enough, before any bifurcation, the system always has two stable equilibria. These correspond to the (QN)-modes (that are CP1). However, the pitchfork bifurcations for the CP1 of the $2:2$ resonance are replaced by saddle-node bifurcations. Since for $\Delta_1\to0$ the bifurcation sequence would coincide with the one of the $2:2$ resonance,  the unstable equilibrium that appears after the saddle-node bifurcation must correspond to the  normal mode of the 2:2 resonance that is unstable after the first pitchfork bifurcation. Thus, this equilibrium becomes stable after the bifurcation of the pair of CP2, via a pitchfork bifurcation that takes place when the red line crosses the ellipse. In fact, since $AB<0$, the CP2 are unstable.
The pair of CP2 disappears then via an anti-pitchfork bifurcation, with one of the quasi-normal modes, that becomes unstable. Then a pair of CP1 disappears with a saddle-node bifurcation, that involves this unstable quasi-normal mode, corresponding to the  unstable normal mode of the $2:2$ resonance in the limit $\Delta_1\to0$.

By increasing the value of $\Delta_1$, the slope $m$ of the straight lines becomes smaller and smaller, as the system moves away from the $2:2$ resonance, and therefore the bifurcation sequence can change further. See the orange and yellow lines, in the first top frame of Figure \ref{fig:CatMapD1p}, where we see that the pitchfork bifurcations first disappear and then the saddle-node bifurcations disappear, respectively. 

However, notice that in our assumptions (given by Eq. \eqref{DeltaAssumptions}), $m\geq1$; thus, the system can move further from the symmetry of the $2:2$ resonance but can never be `closer' to the symmetry at $\beta_3\equiv0$ (Case III) of the `rotated' $2:2$ resonance). In other words, the  line representing a system can not have an angular coefficient smaller than one, and thus its slope can never be closer to the one of the horizontal line at $\beta_3\equiv0$ than to the one of the vertical line at $\beta_1\equiv0$; in the worst case, the system could be at the ``same distance'' from both cases II) and III) (when $m=1$).

\begin{figure}[htpb]
\begin{center}
\includegraphics[scale=.5]{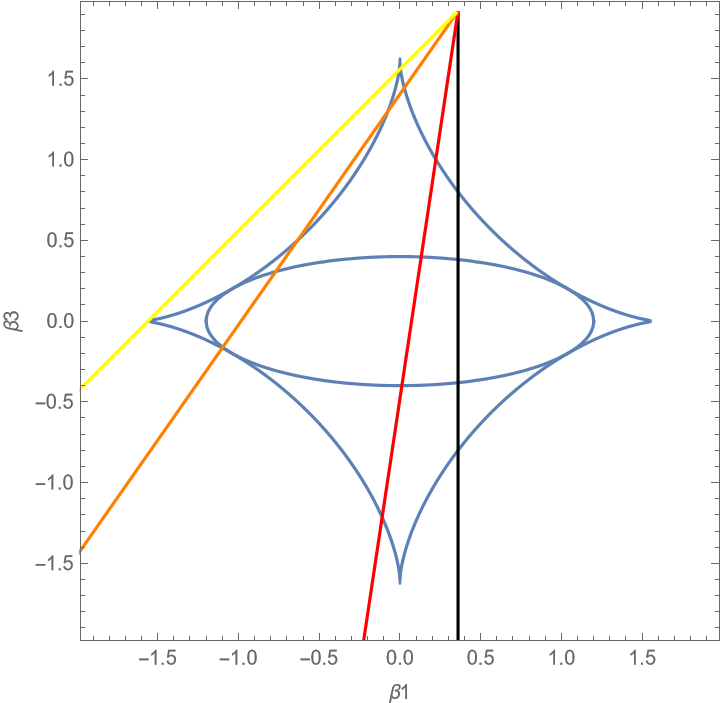}
\quad
\includegraphics[scale=.5]{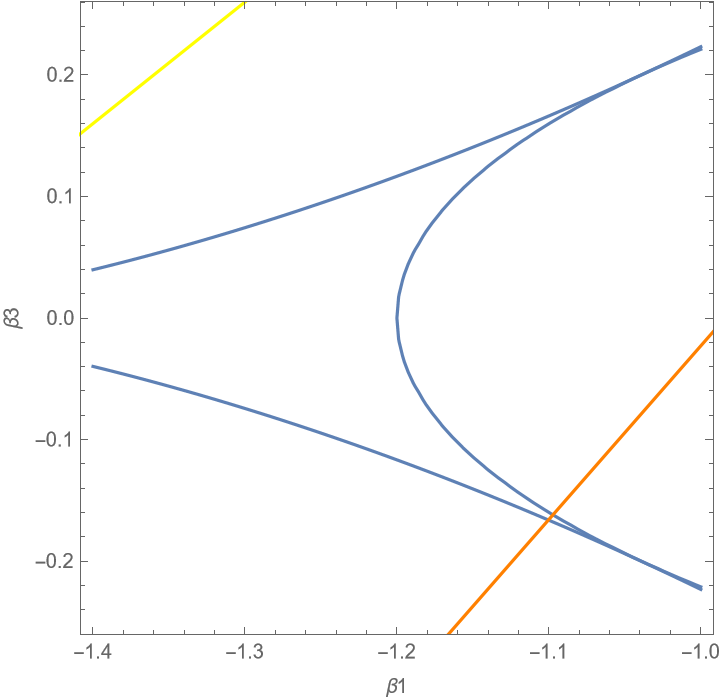}
\quad
\includegraphics[scale=.5]{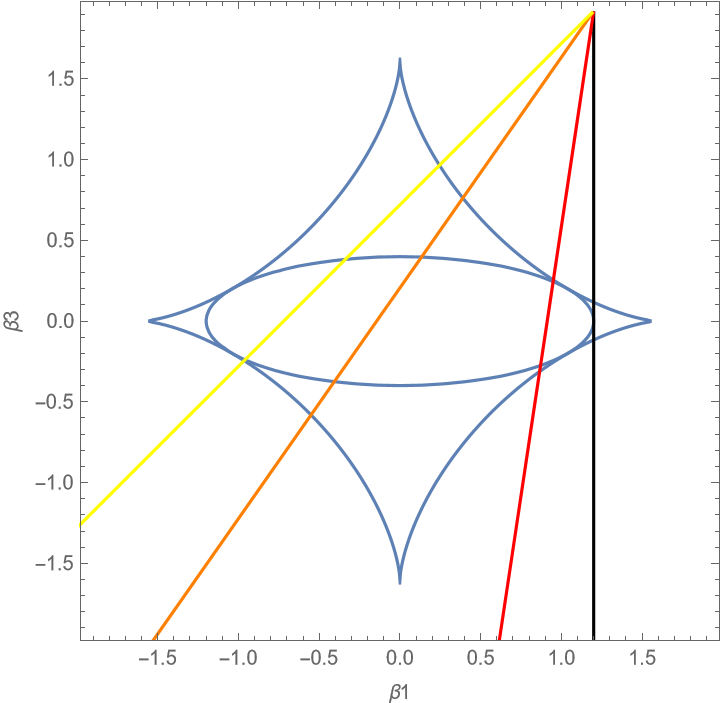}
\quad
\includegraphics[scale=.5]{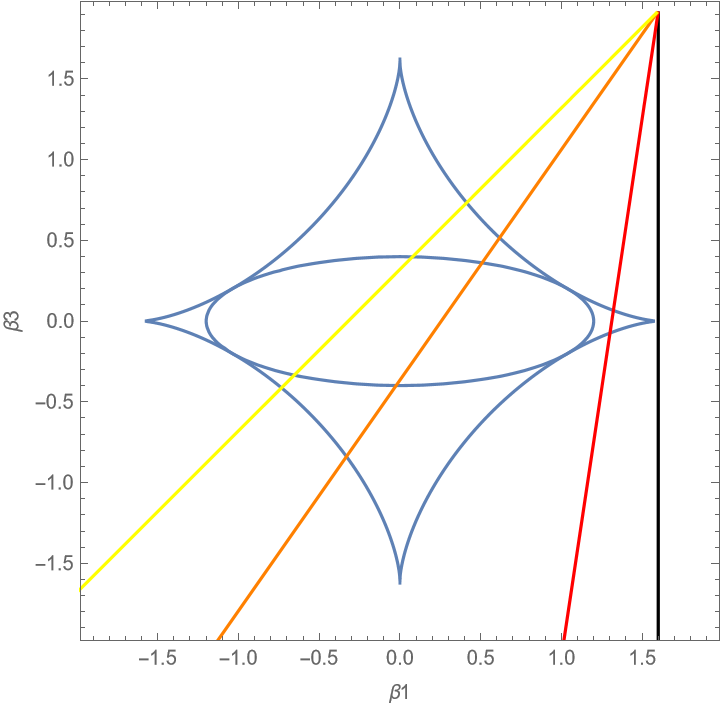}
\end{center}
\caption{Catastrophe maps for $\alpha=-\frac35$ ($A=1, B=-\frac13$), $D_3=-3.2$, $\Delta_3=0.2$. First row: $D_1=-0.6$. In the second panel we zoomed to see the two intersections of the orange line when $\beta_3<0$. Thus, the orange line in the left top frame has a total of six intersection points with the bifurcation curves, showing the possibility of a sequence with six bifurcations.  Second row, from left to right, $D_1=-2$, $D_1=-\frac83$. The straight lines represent different systems at different values of $\Delta_1$: $\Delta_1=0$ (black line), $\Delta_1=0.03$ (red line), $\Delta_1=0.14$ (orange line), $\Delta_1=0.2$ (yellow line).} \label{fig:CatMapD1n}
\end{figure}

If also $D_1\neq 0$, the sequence can be further  away from the one of the  $2:2$ resonance, even for $\Delta_1=0$ (see both Figures \ref{fig:CatMapD1p} and \ref{fig:CatMapD1n}). In particular, we observe that for $D_1>0$, the values of $D_1$ at which the black line ($\Delta_1=0$) touches the ellipse and the star-shaped curve at only one point delimit the appearance/disappearance of pitchfork and saddle-node bifurcations, respectively. For $D_1<0$, even sequences with six bifurcations are possible; see Figure \ref{fig:CatMapD1n}, first row.


\begin{figure}[htpb]
\begin{center}
\includegraphics[scale=.5]{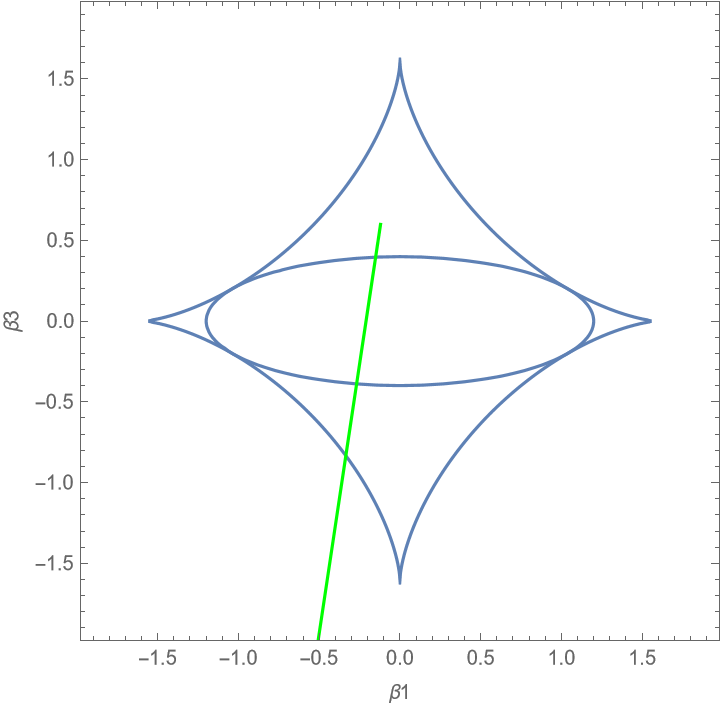}
\end{center}
\caption{Catastrophe map when the parameters are chosen as in Figures \ref{fig:hyperbolas1}-\ref{fig:hyperbolas2} ($A=1, B=-\frac13$, $D_3=-1$, $\Delta_3=0.2$, $D_1=0.2$). Thus $\alpha=-\frac35$.} \label{fig:CatMapFigHyperbola}
\end{figure}

Let us observe once again that, depending on the value of $D_1,D_3$, the bifurcation sequence could end earlier, as  the coordinates of $P_{\infty}$ also depend on $D_1$ and $D_3$ (compare  Figure \ref{fig:CatMapFigHyperbola} and Figure \ref{fig:CatMapD1p}, second frame). The systems represented by the green and red lines differ only in the choice of $D_3$. In the first case, the green line crosses the bifurcation curves three times, thus the system passes through three bifurcations. In the second case, the red line has a fourth intersection, corresponding to an additional fourth bifurcation for the system. See also Figures \ref{fig:hyperbolas1} and \ref{fig:hyperbolas2} (here a change of $D_1$ does not affect the sequence of Figure \ref{fig:hyperbolas1}).

\subsubsection{$\alpha<-1$}
\begin{figure}[htpb]
\begin{center}
\includegraphics[scale=.5]{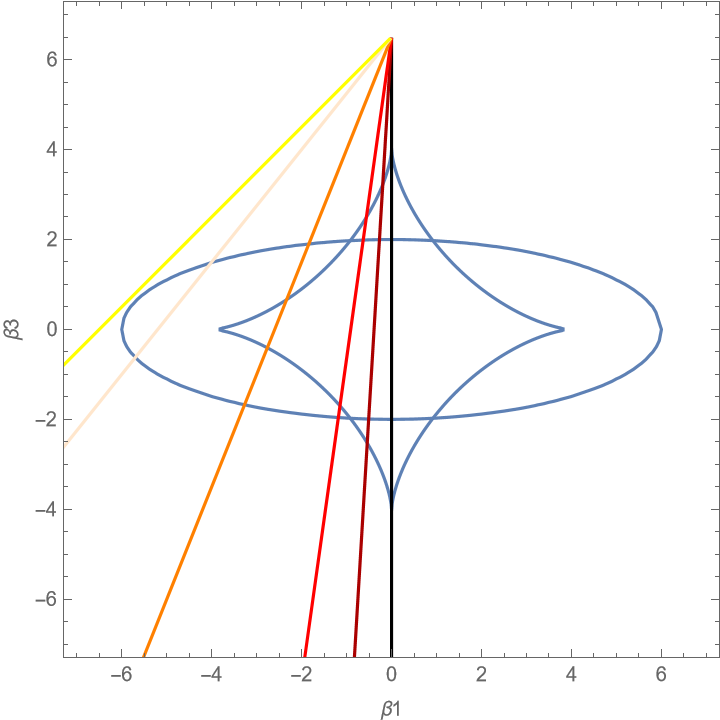}
\end{center}
\caption{Catastrophe map when the parameters are chosen as $A=3$, $B=1$ (thus, $\alpha=-3$),  $D_1=0$, $D_3=-6.5$, $\Delta_3=0.5$ at different values of $\Delta1$: $\Delta_1=0$ (black line), $\Delta_1=0.03$ (brown line), $\Delta_1=0.07$ (red line), $\Delta_1=0.2$ (orange line), $\Delta_1=0.4$ (pink line), $\Delta_1=0.5$ (yellow line).  By keeping $\Delta_3$ fixed and increasing the value of $\Delta_1$ the slope of the lines decreases to $1$, showing all possible bifurcations sequences when $D_1=0$. } \label{fig:CatMapAlpham1}
\end{figure}

\begin{figure}[htpb]
\begin{center}
\includegraphics[scale=.4]{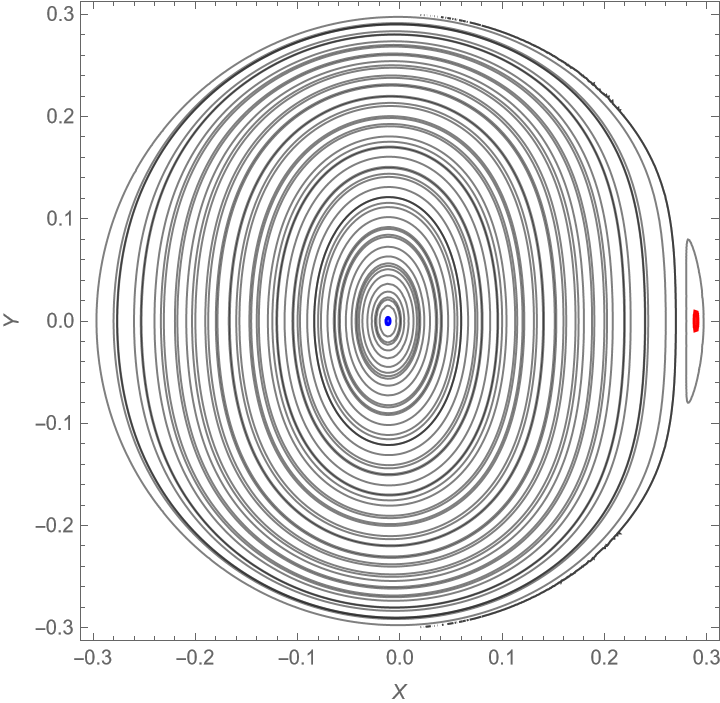}
\quad
\includegraphics[scale=.4]{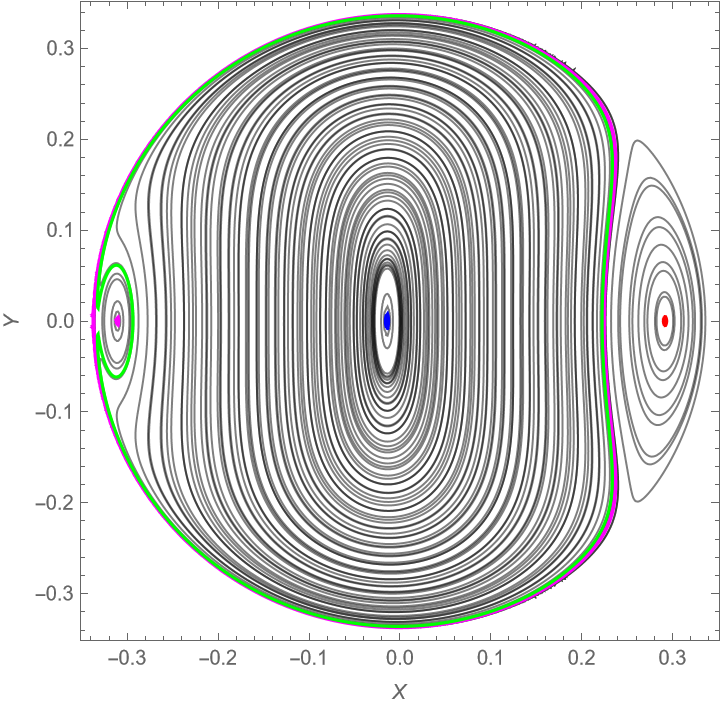}
\quad
\includegraphics[scale=.4]{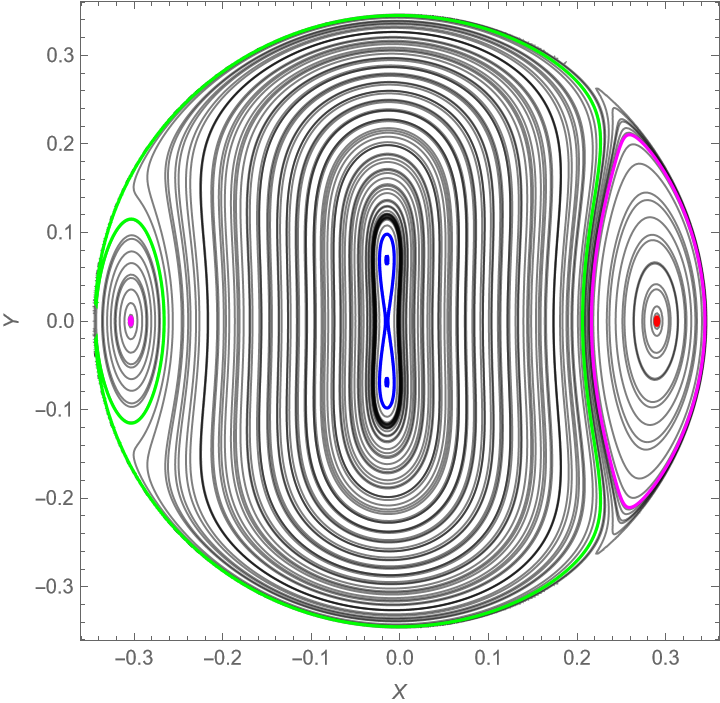}
\quad
\includegraphics[scale=.4]{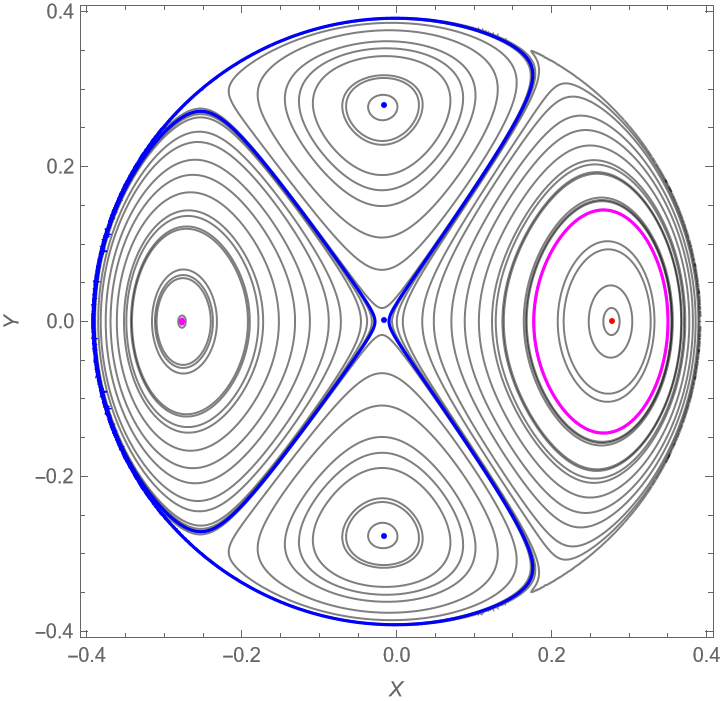}
\quad
\includegraphics[scale=.4]{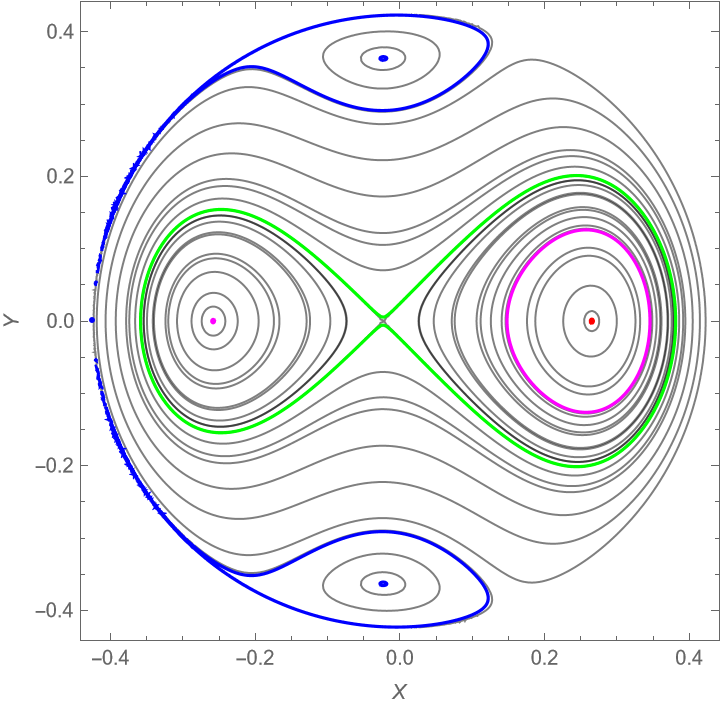}
\quad
\includegraphics[scale=.4]{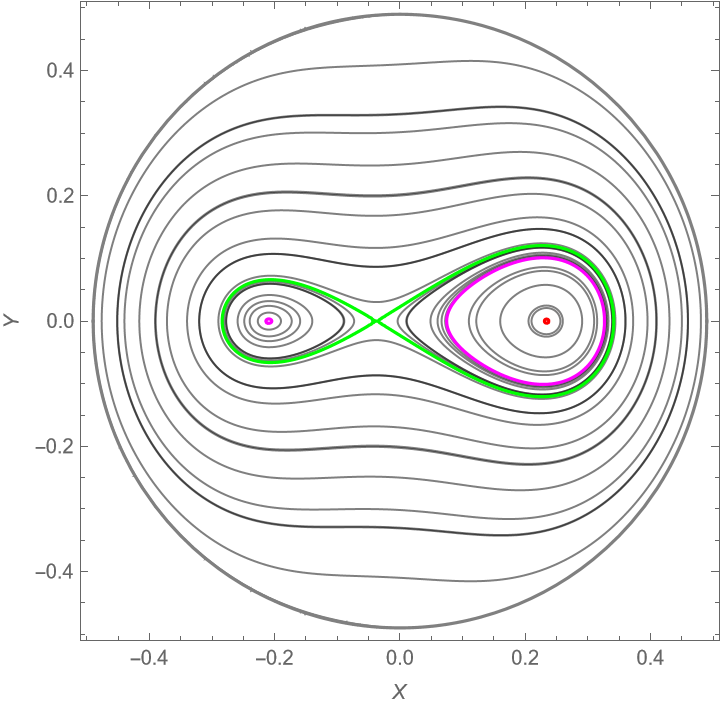}
\quad
\includegraphics[scale=.4]{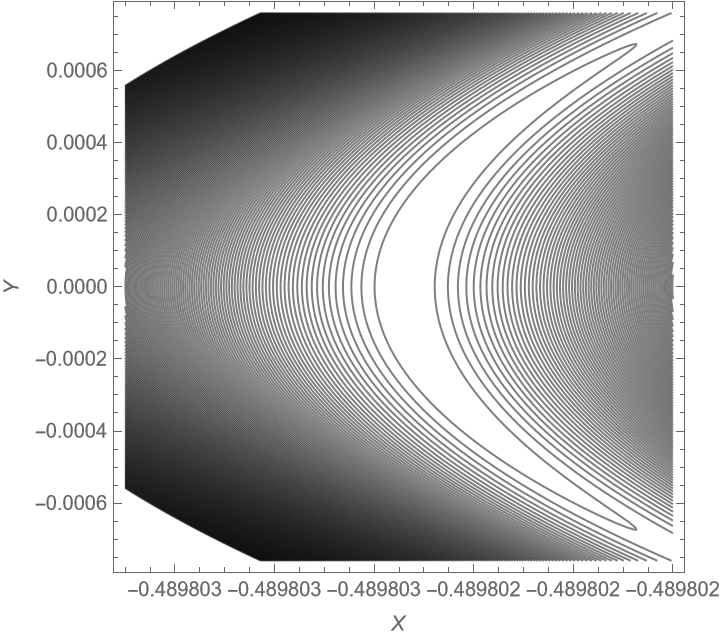}
\quad
\includegraphics[scale=.4]{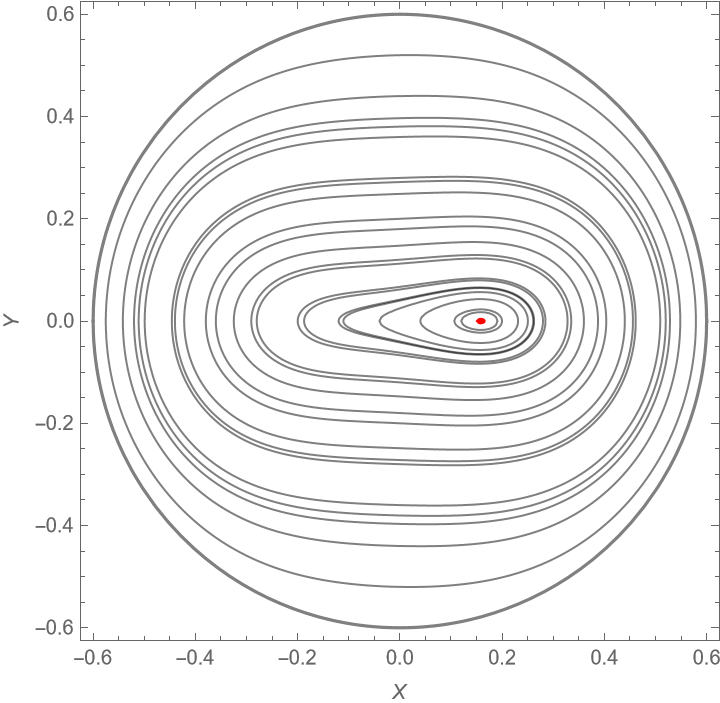}
\end{center}
\caption{Surfaces of section in the $(x,y)$ stereographic coordinates \eqref{ccoord}, illustrating the bifurcation sequence for $\alpha=-3$ corresponding to the brown line of Figure \ref{fig:CatMapAlpham1}. Top row, from left to right: $\E=0.045$, $\E=0.057$, $\E=0.06$. Middle row, from left to right $\E=\E_{gb,1}=0.077$, $\E=0.09$, $\E=0.12$. Bottom left, $\E=0.12$: zoom to show  an equilbrium so close to the left boundary of the Poincar\'e surface that it is not visible in the previous frame. Bottom right: $\E=0.18$, besides the equilibrium in red, also in this case there is another stable equilibrium that is too close to the boundary of the Poncar\'e surface to be seen. } \label{fig:Alpham1PS}
\end{figure}

Let us here take $\alpha$ still negative, but smaller than $-1$. In this case, in the $(\beta_1, \beta_3)$ plane the curves of the bifurcations of the CP2 are oblate ellipses that intersect the curves \eqref{eqBifCP1AlphaBeta} of the CP1 in $4$ points (see Figure \ref{fig:BifCurves2D}, left). Taking $D_1=0$, we see that there can be up to six possible bifurcation sequences, as illustrated in Figure \ref{fig:CatMapAlpham1}. In black we have the vertical line of the $2:2$ resonance. Moving towards the point at infinity, the black line first intersects the star-shaped curve: a pair of inclined orbits (CP1) appears from one normal mode via a pitchfork bifurcation. Since $\alpha<-1$, we have $|A|>|B|$, so these CP1 are stable (see Section \ref{sec:2:2}). The black line then intersects the ellipse: a pair of CP2 (loop orbits) appears via a pitchfork bifurcation. Since $AB>0$, the CP2 are  stable, too. Then the bifurcation sequence continues with two anti-pitchfork bifurcations that determine first the disappearance of the CP2 and then of the CP1. By increasing the value of $\Delta_1$, we obtain the possible sequences of the generic case IV). Moving along  the brown line, from bottom to top, the sequence follows as the one of the $2:2$ resonance, with the difference that  the (anti-)pitchfork bifurcation of the CP1 are replaced by  saddle node bifurcations. The unstable equilibrium that appears  after the saddle node bifurcation corresponds to the normal mode of the 2:2 resonance that is unstable  after the pitchfork bifurcation. See Figure \ref{fig:Alpham1PS} for the corresponding surfaces of section. 

Both  bifurcating CP1 and CP2 are stable for the $2:2$ resonance, which therefore has a global bifurcation (\cite{bif11}) when crossing the symmetry at $\beta_3=\delta_3=0$. This phenomenon can also be observed in the generic case. See the first panel, second row of Figure \ref{fig:Alpham1PS}, which illustrates the global bifurcation corresponding to  Figure \ref{fig:CatMapAlpham1} when the brown line crosses the $\beta_3$ axis. We expect the global bifurcation to always appear in the generic cases for the values of $\alpha$ that make it possible for the $2:2$ resonance. However, a  rigorous proof is beyond the scope of this work.

The intersection points of the two bifurcation curves determine an exchange of the bifurcation order between CP1 and CP2, as we can see, for instance, looking at the red and orange lines in Figure \ref{fig:CatMapAlpham1}. For a greater value of $\Delta_1$ we observe the disappearance of the saddle node bifurcations (pink line), followed by the disappearance of all bifurcations (yellow line). 

Figure \ref{fig:CatMapAlpham1} shows a case in which the bifurcation sequences have the maximal number of steps. However, depending on  $D_3$, the point at infinity can move along the direction of $\beta_3$, thus, in principle, a bifurcation sequence could terminate earlier. 

For $D_1\neq0$, the point at infinity moves to the left ($D_1<0$) or to the right ($D_1>0$), causing the system to move even further from the $2:2$ resonance, and, correspondingly, the possible bifurcation scenarios can change further, in the same fashion as in Figures \ref{fig:CatMapD1p}, \ref{fig:CatMapD1n} of the previous case.

\subsubsection{$0<\alpha<1$}
In case $0<\alpha<1$ the ellipses of the bifurcations of the CP2 are prolate and intersect the star-shaped curves of the CP1 at $4$ points for $0<\alpha<\frac13$. For $\alpha=\frac13$ the two curves intersect only at the two vertices of the star-shaped curves on the $\beta_1$ axis. Otherwise, for $\frac 13<\alpha<1$ the two curves have no contact points. See Figure \ref{fig:BifCurves2D} (right) for an illustration. 
\begin{figure}[htpb]
\begin{center}
\includegraphics[scale=.4]{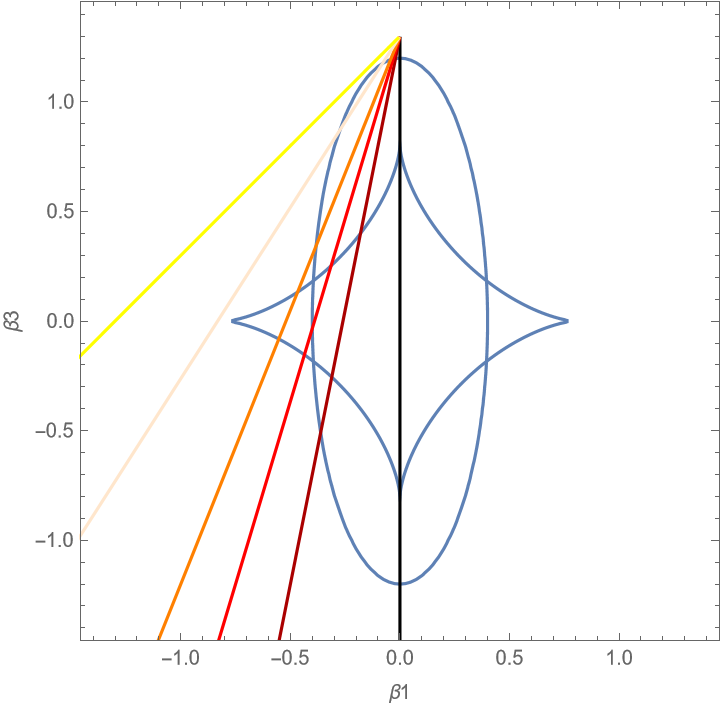}
\quad
\includegraphics[scale=.4]{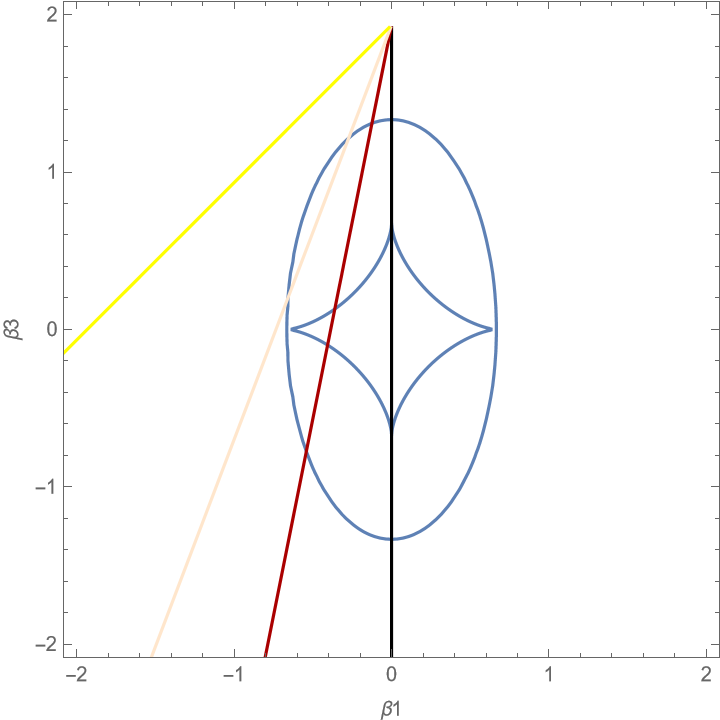}
\quad
\includegraphics[scale=.4]{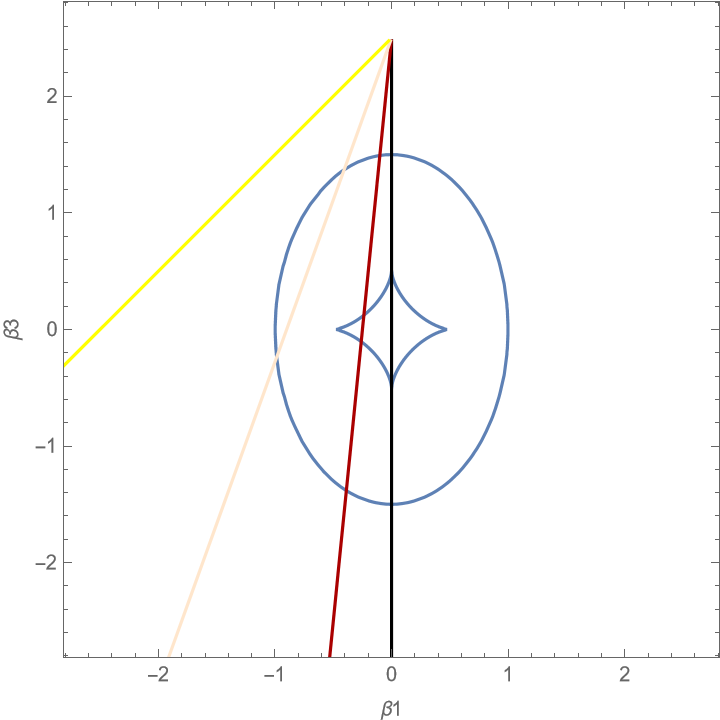}
\end{center}
\caption{Catastrophe maps for $\alpha$ chosen so that $\alpha<\frac13$ (left), $\alpha=\frac13$ (middle) and $\alpha>\frac13$ (right) and $D_1=0$. Namely,  on the left the parameters of the system take values $B= -3$, $A=-1$ (thus, $\alpha=\frac15$, $D_1=0$, $D_3=-6.5$,  $\Delta_3=0.5$
at different values of $\Delta1$: $\Delta_1=0$ (black line), $\Delta_1=0.1$ (brown line), $\Delta_1=0.15$ (red line), $\Delta_1=0.2$ (orange line), $\Delta_1=0.31$ (pink line), $\Delta_1=0.5$ (yellow line). 
In the middle we have $B= -2$, $A=-1$ (thus, $\alpha=\frac13$, $D_1=0$, $D_3=-5.8$, $\Delta_3=0.5$
at different values of $\Delta1$: $\Delta_1=0$ (black line), $\Delta_1=0.1$ (brown line),  $\Delta_1=0.172$ (pink line), $\Delta_1=0.5$ (yellow line).
On the right  the parameters  take values $B= -1.5$, $A=-1$ (thus, $\alpha=\frac12$, $D_1=0$, $D_3=-5$, $\Delta_3=0.5$
at different values of $\Delta1$: $\Delta_1=0$ (black line), $\Delta_1=0.05$ (brown line),  $\Delta_1=0.18$ (pink line), $\Delta_1=0.5$ (yellow line). } \label{fig:CatMapAlpha01}
\end{figure}
These differences in the contact points of the two curves do not affect the bifurcation sequence of the $2:2$ resonance; see the black line in Figures \ref{fig:CatMapAlpha01}. The bifurcation sequence always starts with a pair of CP2 that appear from one of the normal modes via a pitchfork bifurcation. Since $0<\alpha<1$ implies $AB>0$, these CP2 are stable. Once the black line meets the star-shaped figure, we have the appearance of a pair of CP1 via another pitchfork bifurcation. These CP1 are unstable since for $0<\alpha<1$ we have $|A|<|B|$. Therefore, they must appear from the same normal mode  the CP2 appeared from, since this was the only orbit unstable before the bifurcation of the CP1. Then the sequence continues with the other two anti-pitchfork bifurcations from the other normal mode (see also \cite{bif11}). 
From Figure \ref{fig:CatMapAlpha01} we see that for $D_1=0$ and $\Delta_1$ small enough, in the generic case  the sequence closest to the $2:2$ resonance remains unchanged, independently on $0<\alpha<1$. Namely, in all three panels of Figure \ref{fig:CatMapAlpha01} the sequences corresponding to the brown lines are the same. Below any critical value for $\E$, there are only two stable equilibria, corresponding to the (QN)-modes. By increasing the energy, we move on the brown line towards the point at infinity. Once the brown line intersects the ellipse, we have a pitchfork bifurcation that determines the appearance of a pair of stable CP2 from one of the (QN)-modes, that becomes unstable. Afterwards, at the crossing points with the star-shaped figure, there are the saddle-node bifurcations. The unstable CP1 that appear/disappear via the saddle-node bifurcation corresponds to one of the unstable CP1 that appeared via the pitchfork bifurcation for the $2:2$ resonance. Finally, there is an anti-pitchfork bifurcation, that determines the disappearance of the stable CP2 from the unstable (QN)-mode. 

However, for $\alpha<\frac13$, the intersection points between the two bifurcation curves  determine two possible exchanges between the saddle-node and the pitchfork bifurcations in the sequence for the generic case of the inversion symmetry; see, for instance, the red and orange lines in Figure \ref{fig:CatMapAlpha01} (left).  This does not happen for $\alpha\geq\frac13$. In particular, for $\alpha=\frac13$ the two curves intersect at the tip points of the star- shaped figure when $\beta_3=0$. As discussed in Section \ref{sec:catmapdelta3zero}, this correspond to a structurally unstable case.

A sequence with only the (anti-)pitchfork bifurcations of the CP2 is always possible (see the pink line in Figure \ref{fig:CatMapAlpha01}). Depending on $D_1\neq0$, less/more bifurcation scenarios could be possible. 

In this case a global bifurcation for the $2:2$ resonance  does not take place when crossing the symmetry at $\beta_3=0$, and we expect this to remain true also for the generic case.

\subsubsection{$\alpha>1$}
\begin{figure}[htpb]
\begin{center}
\includegraphics[scale=.5]{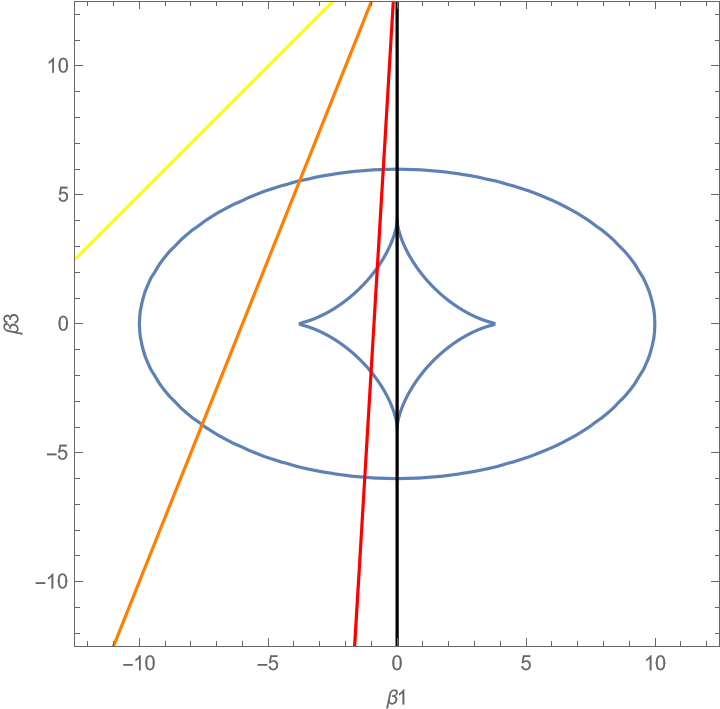}
\quad
\includegraphics[scale=.5]{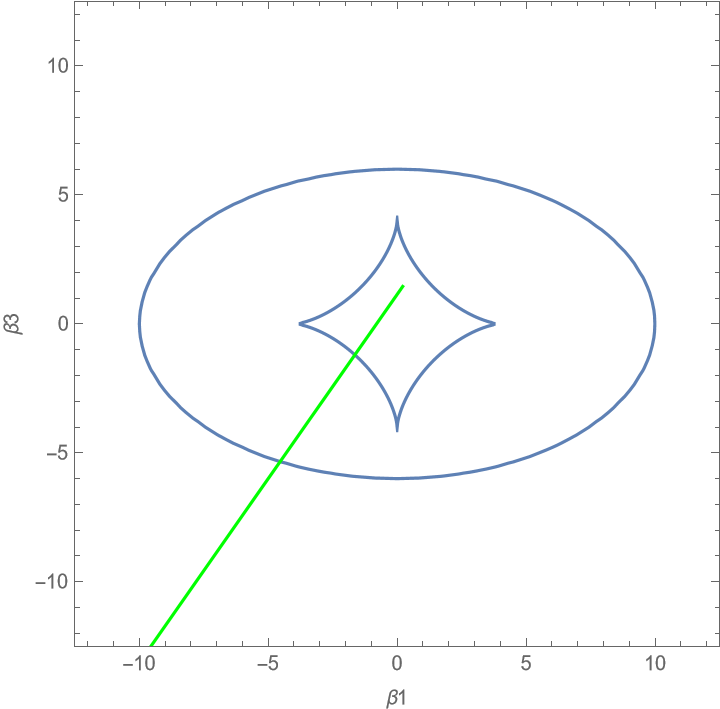}

\end{center}
\caption{Catastrophe maps for $\alpha$ chosen so that $\alpha>1$.  On the left the parameters of the system take values $B= -.6$, $A=-1$ (thus, $\alpha=5$), $D_1=0$, $D_3=--3$,  $\Delta_3=0.5$
at different values of $\Delta1$: $\Delta_1=0$ (black line), $\Delta_1=0.03$ (red line), $\Delta_1=0.2$ (orange line), $\Delta_1=0.5$ (yellow line), 
On the right, after proper transformations, the parameters correspond to the ones of Figures \ref{fig:nosymmgb} and \ref{fig:nosymmgbps}. Namely,  $B= -.6$, $A=-1$ (thus, $\alpha=5$), $\Delta_1=0.07$, $D_1=-0.05$, $D_3=-0.3$, $\Delta_3=0.1$. The bifurcation sequence in this case ends earlier and agrees  with the one showed by Figures \ref{fig:nosymmgb} and \ref{fig:nosymmgbps}.} \label{fig:CatMapAlpha0M1}
\end{figure}

Let us conclude with the case $\alpha>1$. The bifurcation curves of CP2 are oblate ellipses in the $(\beta_1, \beta_3)$ plane that always surround the star-shaped curve of CP1, without contact points. As previously, let us focus on the case $D_1=0$. To find the bifurcation sequence of the $2:2$ resonance, we look at Figure \ref{fig:CatMapAlpha0M1} and  move on the black line towards the point at infinity. Therefore, we see that the sequence starts with a pitchfork bifurcation for the CP2. Since $\alpha>1$ implies $AB>0$, the CP2 are stable. The sequence then continues with the appearance of a pair of CP1, which also are stable, since in this case $|A|>|B|$. Since these CP1 are stable, they correspond to the bifurcation of a pair of stable inclined orbits, which do not appear from the same normal mode as the loop orbits (corresponding to CP2), as this is already unstable after the first pitchfork bifurcation. Since both normal modes are now unstable, we have a global bifurcation when the black line crosses the $\beta_1$ axis \cite{bif11}. Afterwards, we have two anti-pitchfork bifurcations that determine the disappearance of first the CP2 and then the CP1. In the generic case of the red line, the bifurcation sequence follows similarly, with the pitchfork for the CP1 replaced by a saddle-node bifurcation. Moving further from the symmetry of the $2:2$ resonance we observe first the disappearance of the pair of saddle-node bifurcations (orange line) and then the disappearance of the pitchfork and anti-pitchfork bifurcation (yellow line).

In all the cases studied, we assumed the condition \eqref{DeltaAssumptions}. Let us briefly show how easily one can deduce the bifurcation sequence if the parameters do not satisfy \eqref{DeltaAssumptions}. Let us take as a demonstrative example the parameter choice of Figure \ref{fig:nosymmgbps} (or equivalently, Figure \ref{fig:nosymmgb}) where $\Delta_1=-0.1$ and $\Delta_3=-0.07$ do not satisfy \eqref{DeltaAssumptions}. We can apply the transformations $I_1\longrightarrow -I_1$ and $I_3\longrightarrow -I_3$ to Hamiltonian \eqref{hamiltonian_inv_noC} to achieve a new system for which $\Delta_1, \Delta_3 \geq0$ (and also $D_1$, $D_3$ have to change sign). Then, by exchanging $I_1$ and $I_3$, we obtain a system with $\Delta_1=0.07$ and $\Delta_3=0.1$, satisfying \eqref{DeltaAssumptions}. Moreover, $A$ exchanges with $B$, so now $A=-1$, $B=-0.6$. Thus, we are in case $\alpha>1$. The corresponding catastrophe map is shown in Figure \ref{fig:CatMapAlpha0M1} and it is in agreement with the bifurcation sequence of the original system (see Figure \ref{fig:nosymmgbps}). Notice the appearance of a global bifurcation when the system crosses the symmetry at $\delta_1=\beta_1=0$, as for  the case of the $2:2$ resonance. 

\section{Conclusions}\label{sec:conclusions}
We have studied the bifurcations  of the detuned 1:1 Hamiltonian resonance with inversion symmetry. The normal form for the general 1:1 resonance is much more complicated than the one considered in this work \cite{Ku, CushmanRod,Dui, HH2019}. However, symmetry often naturally appears in applications, with the advantage of considerably simplifying the normal form.

As already observed, the inversion symmetry \eqref{inv_simm} we consider here emerged after averaging in the three body problem \cite{RC23, RAG25}. The 2:2 resonance, with the reflectional symmetries $\mathcal S_1$, $\mathcal S_2$, cf.\eqref{spatial_symmetry_1}, \eqref{spatial_symmetry_2}, appears in galactic dynamics to describe the orbital structure of elliptical galaxies \cite{11-2011,MNRAS}.  In this work we see how, by breaking the mirror symmetries $\mathcal S_1$, $\mathcal S_2$, cf.\eqref{spatial_symmetry_1}, \eqref{spatial_symmetry_2}, of the 2:2 resonance, but keeping the symmetry under $\mathcal S_1\oplus\mathcal S_2$, one of the pitchfork bifurcations of the 2:2 resonance turns into a saddle-node bifurcation. 

As long as the equilibria of the reduced system on the Hopf sphere, cf. \eqref{rel_invariants}, are isolated, they correspond to periodic orbits for the dynamics in two degrees of freedom.  A pair of isolated equilibria always exists. These are CP1, which are rectilinear segments in the $(Q_1, Q_2)$ configuration plane, corresponding to the quasi-normal modes. For values of $\E$ below any bifurcation threshold, these quasi-normal modes become  the normal modes of the 2:2 resonance (orthogonal rectilinear segments) in the limit for $\delta_1\to0$.  

In Section \ref{sec:bifurcations} we see that the Hamiltonian \eqref{hamiltonian_inv_noC} can have up to four additional equilibria on the Hopf sphere: a pair of CP1 that always appear through a  saddle-node bifurcation and a pair of CP2 that always bifurcates via a pitchfork bifurcation.

In the case of the 2:2 resonance, both CP1 and CP2 appear via pitchfork bifurcations and correspond to the so called inclined and loop orbits for the system in  two degrees of freedom, determined by the conditions $I_2=0$ and $I_1=0$, respectively. Thus, inclined orbits are rectilinear segments related to oscillations of both degrees of freedom in phase ($\psi=0,\pi$), while loop orbits are closed orbits related to oscillations of both degrees of freedom in quadrature ($\psi=\pm \pi/2$). 

In the case of inversion symmetry, the CP2 are not constrained anymore on the plane $I_1=0$.
The value of the constant phase angle $\psi$ of the corresponding periodic orbits, cfr. \eqref{suber}, can be easily found from the coordinates \eqref{EquiCenter} of the equilibria on the Hopf sphere, namely
$$ \cot\psi= \frac{\mathcal {Q}_1}{\mathcal {Q}_2}=\pm \frac{\delta_1 A B}{\sqrt{2(A^2 B^2 \E^2-\delta_1^2 B^2-\delta_3^2 A^2})}\,. $$
For $\delta_1=0$, we recover the condition $\psi=\pm \pi/2$ of the 2:2 resonance.
The CP1 still lie on the $I_2$ plane, thus they could still be understood as inclined orbits.
However, in the limit $\delta_1\to 0$, when the system reaches the symmetry of the 2:2 resonance, the CP1 of the inversion symmetry can become either inclined orbits or normal modes (orthogonal rectilinear segments, not associated to a condition on the phase angle $\psi$). 

We provide analytical formulas that determine the bifurcations, cf. \eqref{eqBifCP2}, \eqref{eqBifCP1}. Once the value of the detunings $\Delta_1, \Delta_3$ and of the coefficients of the normal form are known for a specific system,  equations \eqref{eqBifCP2}, \eqref{eqBifCP1} can be solved numerically, providing critical values for the distinguished parameters $\E$ (the radius of the Hopf sphere) that determine the bifurcations, see Section \ref{sec:bifurcations}.
These can be used to predict the bifurcations of the original system. Once  the critical values of $\E$ are calculated, they do not provide a complete understanding of the phase space structure before and after each such threshold values. The bifurcation scenario is manifold, with bifurcation sequences that could display the alternating of pitchfork and saddle-node bifurcations. The pitchfork bifurcation could take place from one of the equilibria appeared after the saddle-node bifurcation, or it could be completely unrelated to it.
In Section \ref{sec:catamap} we find, via  the catastrophe map, an effective way to discern the possible bifurcation sequences and predict  the phase space scenario  depending on the value of the coupling parameter $\alpha$. The inspiration to use the catastrophe map came to us from the chemical literature, where it was implemented in the study of molecular systems \cite{VK,STK}. On the other hand, the most direct application is to the planetary three-body system, in which the periodic orbits in generic position correspond to configurations in which the quasi-Keplerian orbits of the two `planets' have fixed mutual orientation (corotation or anti-corotation of the apsides) and the role of the Hopf radius id played by the {\it angular momentum deficit} \cite{RC23, RAG25}.

\section*{Acknowledgments}
We are grateful to F. Verhulst for very fruitful discussions and to the Organizers of the SPT Conference Series for the inspiring atmosphere where most of these conversations took place.

A.M. and G.P. are members of GNFM-INdAM. 

A.M. is supported by the Ministry of Education, Youth and Sports of the Czech Republic within the institutional support for long-term development of a research organization at Czech Technical University in Prague (RVO 68407700).

\end{document}